\documentclass[12pt,a4paper]{article}
\pdfoutput=1 % if your are submitting a pdflatex (i.e. if you have
\usepackage{jheppub} % for details on the use of the package, please
\usepackage{amsmath,amssymb,amsthm,amscd}
\usepackage{dsdshorthand}
\usepackage{mathtools} 
\usepackage{bbm}
\usepackage{arydshln}
\newcommand{\tr}{{\rm Tr}}
\newcommand{\beq}{\begin{equation}}
\newcommand{\eeq}{\end{equation}}
\usepackage[T1]{fontenc}

\renewcommand{\be}{\begin{eqnarray}}
\renewcommand{\ee}{\end{eqnarray}}
\newcommand{\bea}{\begin{eqnarray}}
\newcommand{\eea}{\end{eqnarray}}

\def\XXint#1#2#3{{\setbox0=\hbox{$#1{#2#3}{\int}$}
     \vcenter{\hbox{$#2#3$}}\kern-.5\wd0}}

\newcommand{\remarkjc}[1]{{\renewcommand{\bfdefault}{b}{\color[RGB]{0,0,150}{\textbf{#1}}}}}
\providecommand{\remarkjc}[1]{\ignorespaces}

\title{Holography for $\mathcal{N}=4$  on $\mathbb{RP}^4$ 
}

\author[\hspace{.08em}a,b,c]{Jo\~ao~Caetano}
\author[\hspace{.08em}b]{, Leonardo Rastelli}

\affiliation[a]{Department of Theoretical Physics, CERN, 1211 Meyrin, Switzerland
}
\affiliation[b]{C. N. Yang Institute for Theoretical Physics, Stony Brook University, Stony Brook, NY 11794, USA}
\affiliation[c]{Simons Center for Geometry and Physics, Stony Brook University, Stony Brook, NY 11794, USA }

\preprint{CERN-TH-2022-097}

\abstract{
We propose a holographic description of $\mathcal{N}=4$ super Yang-Mills on the four-dimensional real projective space $\mathbb{RP}^4$. 
We first construct the dual background in the framework of  five-dimensional $\mathcal{N}=8$ gauged supergravity, and then uplift  it to a new one-half BPS solution of type IIB supergravity. A salient feature of our solution is the presence of a bulk naked singularity whose 
local 
 behavior resembles that of an O1$_{-}$ plane in flat space. 
}

\begin{document}
\maketitle

\flushbottom

\section{Introduction}

In this paper we study the holographic description of $\mathcal{N}=4$ super Yang-Mills with SU($N$)  gauge group  on the four-dimensional real projective space $\mathbb{RP}^4$,
the simplest unorientable  four-manifold.  
There are several reasons to consider this generalization of the paradigmatic AdS/CFT duality. 
Any quantum field theory with  time reversal symmetry can be defined on an unorientable spacetime. This construction has been developed both 
in the theoretical condensed matter  \cite{Wang:2003cu, Hsieh:2015xaa, Metlitski:2015yqa, Barkeshli:2016mew, Guo:2017xex, Wan:2018zql, Wan:2019oyr, Wang:2019obe, Wang:2020jgh} and high-energy   \cite{Kapustin:2014tfa, Kapustin:2014gma, Kapustin:2014dxa,  Witten:2015aba, Seiberg:2016rsg,  Witten:2016cio, Tachikawa:2016cha, Tachikawa:2016nmo} literature, partly in order  to probe certain subtle  anomalies involving time reversal.
It is of interest to study such a setup at strong coupling. The holographic duality for  $\mathcal{N}=4$ SYM offers a direct window into  strongly coupled dynamics that can serve as a simplified model for more general QFTs.\footnote{Holography for field theories on non-orientable manifolds in two dimensions has been considered in \cite{Maloney:2016gsg}. Crosscap states have also been studied in the context of the bulk reconstruction program in holography, as possible duals to fields inserted at a bulk point, see e.g.~\cite{Miyaji:2015fia, Verlinde:2015qfa,  Nakayama:2015mva,  Nakayama:2016xvw, Goto:2016wme, Lewkowycz:2016ukf}.}
Another motivation comes from the conformal bootstrap. Formulating a $d$-dimensional CFT  on the real projective space  breaks the (Euclidean) conformal group $SO(d+1, 1)$ to the isometry group $SO(d+1)$ of $\mathbb{RP}^d$. A consequence of this conformal symmetry breaking is the appearance of a new set of observables, namely the one-point functions of scalar operators~\cite{ Nakayama:2016cim,  Hasegawa:2016piv, Hasegawa:2018yqg, Hogervorst:2017kbj, Giombi:2020xah}, which must satisfy certain bootstrap constraints \cite{Giombi:2020xah}.
 This is in some ways analogous to considering CFT in a spacetime with a boundary that preserves a subgroup $SO(d, 1)$
of the full conformal group, but it is a more rigid construction, as there appears to be much less freedom in formulating the CFT on  $\mathbb{RP}^d$ than there is in choosing a consistent boundary state. 
One is curious about this bootstrap problem for  $\mathcal{N}=4$ SYM, the canonical example of a four-dimensional CFT, and especially about its interplay with planar integrability.

The real projective space $\mathbb{RP}^4$ is the compact  unorientable manifold  obtained by modding out the four dimensional sphere $S^4$ by the involution that 
identifies antipodal points. 
To formulate ${\cal N}=4$ SYM  on $\mathbb{RP}^4$ we need to specify how this involution 
acts on the elementary fields.
We will focus on a choice
that preserves the maximal amount possible of supersymmetry, namely 16 of the original 32 supercharges. One has the additional freedom of including in the definition of the involution 
the discrete symmetry of charge conjugation,
which acts by complex conjugation of the SU($N$) gauge group generators. 
There are then (at least) two distinct ways to realize $\mathcal{N}=4$ SYM $\mathbb{RP}^4$ while preserving 16 supercharges. 
They are physically very different. 

If one  includes charge conjugation in the definition of  the involution, 
one-point functions are of order $O(1)$ in the large $N$ limit, while if one does {\it not} include it they are of order $O(N)$. 
In this paper we focus on the latter choice. It is the same choice considered in \cite{Wang:2020jgh}, where  supersymmetric  localization was used to derive a matrix model capturing a protected subsector. Here we endeavor to construct the holographic dual. 

The large $N$ scaling of one-point functions  makes it clear that we need to look for a new classical background, a new one-half BPS solution of IIB supergravity.
The background must be asymptotic to $AdS_5 \times S^5$, since the dual field theory is unchanged in the UV.
What's more, field theory expectations dictate the asymptotic behavior of the bulk supergravity fields. 
Following the blueprint of \cite{Bobev:2020fon}, we first find an analytic solution of the BPS equations with the requisite asymptotic behavior in the appropriate truncation to the five-dimensional  $\mathcal{N}=8$ gauged supergravity (adapted to our Euclidean setup), 
and then uplift it to ten dimensions.  It turns out that our physically motivated boundary conditions allow for a one-parameter family of BPS solutions. This last parameter can in principle be fixed by a more sophisticated matching with field theory expectations (e.g.~using results from supersymmetric localization), but we  leave this for future work. 
A distinct feature of our family of solutions is the existence of a naked singularity, resembling that of an orientifold O1$_{-}$ plane in flat space.

The other possibility (where one includes charge conjugation in the definition of the involution)
will be discussed in a separate article \cite{inprogresscc}. The holographic story is much simpler:
the large $N$ scaling implies that the bulk background is unchanged to leading order, apart from the $\mathbb{Z}_2$ (orientifold) projection. What makes this choice interesting is that planar integrability is preserved \cite{inprogresscc}.

\medskip

The paper is organized as follows. In section \ref{fieldth}, we discuss how to put $\mathcal{N}=4$ SYM on $\mathbb{RP}^4$ and the associated preserved symmetries. In section \ref{holo}, we construct the holographic dual in the relevant five-dimensional supergravity  truncation. In section \ref{uplift}, we provide the full ten-dimensional solution in type IIB supergravity and discuss its main features such as the singularity. In section \ref{sectree}, we comment on the field theoretic computation of one-point functions and integrability. We offer some concluding remarks in section \ref{discussion}. Finally, the appendices contain technical details and peripheral material omitted in the main text.

\section{Field Theory on $\mathbb{RP}^4$} \label{fieldth}
The real projective space  $\mathbb{RP}^4$ is defined as the space of all lines passing through the origin of  $\mathbb{R}^5$. Each line is  specified by a non-zero vector in $\mathbb{R}^5$, which is unique up to scalar multiplication. Then $\mathbb{RP}^4$ is 
the quotient space of $\mathbb{R}^{5}-\{0\}$ under the equivalence relation $v \sim \lambda v$ for any real $\lambda \neq 0$. We can restrict to vectors of unit length, and so $\mathbb{RP}^4$ is also the quotient space 
\beq S^{4} / \{ v\sim -v \}\,,
\eeq that is, the four-dimensional sphere with antipodal points identified. This latter definition is the one we will be mostly using throughout this paper.  

We would like to place a CFT on this manifold. In embedding coordinates $X^{A}=(X^{0},X^{\mu},X^5) \in \mathbb{R}^{1, 5}$, $S^4$ with radius $R$ is specified by  $X^{A}= (R,\Omega^{\mu},\Omega^{5})$ subject to $(\Omega^{\mu})^2+(\Omega^{5})^2=R^2$. The antipodal map 
\beq \label{invol1}
(X^{0},X^{\mu},X^5) \sim (X^{0},-X^{\mu},-X^5)  
\eeq leaves the following combination of generators invariant
\beq
M_{\mu \nu}=J_{\mu\nu}\, ,\quad P_{\mu} - K_{\mu} =- 2 J_{\mu 5}
\eeq
where $J_{AB}= -i \left( X_{A} \frac{\partial}{\partial X^{B}}- X_{B} \frac{\partial}{\partial X^{A}}\right)$ are the $\mathfrak{so}(1,5)$ generators of the (Euclidean) conformal group in four dimensions.
The remaining generators pick up a sign under the antipodal map and therefore the residual conformal symmetry is given by
\beq
\mathfrak{so}(5) \subset \mathfrak{so}(1,5)\,.
\eeq
In particular, the isometries of $S^{4}$ are preserved by the antipodal identification, which will be important when we construct the supergravity solution.

As shown by Yifan Wang~\cite{Wang:2020jgh}, one can place
 $\mathcal{N}=4$ SYM on $\mathbb{RP}^4$ in such a way that sixteen supercharges are preserved in addition to the SO($5$) bosonic symmetry.   Let us review his argument, and in the process spell out our notations.  Supersymmetry breaking follows from the antipodal identification of the Killing spinors of $S^4$. To see this explicitly, it is useful to consider $S^4$ in stereographic coordinates $x^{\mu}$ related to the embedding coordinates in the usual way,
\beq \label{stereo}
x^{\mu} = \frac{\Omega^{\mu}}{R+\Omega^5}\,,\quad ds^2_{S^4} = \omega(x)^2\delta_{\mu\nu}dx^{\mu}dx^{\nu}\, , \quad \omega(x) =\frac{2 R}{1+x^2}\,.
\eeq
In particular, the conformal Killing spinors parametrizing 32 supercharges in flat space
\beq 
\epsilon_{\mathbb{R}^4}(x) = \epsilon_{s}+x^{\mu} \tilde{\Gamma}_{\hat\mu} \epsilon_{c}
\eeq
get mapped to $\epsilon_{S^4}(x) = \omega(x)^{1/2} \epsilon_{\mathbb{R}^4}(x)$. Our notations are as follow. We are using a pair of ten-dimensional gamma matrices $\Gamma_M, \tilde{\Gamma}_N$ that act on positive and negative chirality 16-component spinors respectively. We split their indices into four dimensional spacetime  $\mu=1,\dots,4$ and $\mathfrak{so}(5,1)_{R}$ R-symmetry $I=5,\dots,9,0$. For the spacetime indices, we  distinguish between two types, namely $\Gamma_{\hat{\mu}}$ and $\Gamma_{\mu}$, to denote the flat Euclidean base space and the curved spacetime  respectively, with their relation being $\Gamma_{\hat{\mu}} =\omega(x)\, \delta^{\mu}_{\hat{\mu}}  \, \Gamma_{\mu}$. We do not distinguish the remaining R-symmetry indices between hatted and unhatted, $\Gamma_{\hat{I}}=\Gamma_{I}$. Finally, they satisfy the Clifford algebra $\{\Gamma_{\hat{M}},\tilde{\Gamma}_{\hat{N}}\}=2 \delta_{\hat{M}\hat{N}}$.
Additionally, in the above formula, $\epsilon_{s,c}$ are constant 16-component spinors related to the Poincar\'e and conformal supercharges, respectively.

In these coordinates, the antipodal map becomes the involution 
\beq
\mathcal{I}_{\text{SYM}}:x^{\mu} \mapsto -\frac{x^{\mu}}{x^2}\,, 
\eeq
 which is an inversion composed with a full reflection of all coordinates. Under this involution, one assumes the following ansatz~\cite{Wang:2020jgh} for the transformation of the Killing spinor,\footnote{The action of the involution on the spinors squares to one, and this defines a pin$^{+}$ structure on $\mathbb{RP}^4$. In addition, there is still the freedom of choosing the overall sign of the transformation. We have chosen the plus sign.}
\beq \label{killingtransform}
\epsilon_{S^{4}}(x) \mapsto \frac{i}{|x|} x^{\mu} \tilde\Gamma_{\hat\mu} \,\mathcal{R} \,\epsilon_{S^4}(x') \equiv \tilde{\epsilon}_{S^4}(x')\,,\quad (x')^{\mu} \equiv -\frac{x^{\mu}}{x^2}\,.
\eeq
The above spacetime transformation is the familiar one for the action of an inversion on a spinor of conformal weight $-\frac{1}{2}$. In addition, one allows for a matrix $\mathcal{R}$ acting on $\epsilon_{S^4}$ induced by an R-symmetry outer automorphism.

Invariance of the Killing spinor under the involution, i.e. 
$\epsilon_{S^{4}}(x) =\tilde\epsilon_{S^{4}}(x')$, together with the requirement that $\epsilon \gamma_{\mu} \epsilon$ transforms as a vector as required by superconformal symmetry, constrains $\mathcal{R}$ to be
\beq
\mathcal{R} = - \Gamma_{I}\tilde{\Gamma}_{J}\Gamma_{K}\,,
\eeq
and the supercharges are such that the corresponding constant spinors satisfy
\beq \label{halfBPS}
\epsilon_{c} = -i \mathcal{R} \epsilon_{s}\,.
\eeq
Above, $I,J,K$ is any choice of the $\mathfrak{so}(5,1)_{R}$  indices. For definiteness, we make the choice 
$\mathcal{R} = - \Gamma_{790}$, as in  \cite{Wang:2020jgh}. 

The condition (\ref{halfBPS}) breaks the R-symmetry down to $\mathfrak{so}(3)_{568} \times \mathfrak{so}(2,1)_{790}$ and halves the number of supercharges. The preserved  superalgebra is
\beq \label{symm}
\mathfrak{osp}(4^{*}|4) \supset \mathfrak{so}(5) \times \mathfrak{so}(3)_{568} \times \mathfrak{so}(2,1)_{790}\,.
\eeq

In order to fully define the theory on $\mathbb{RP}^4$ we must specify the transformations of the elementary SYM fields.
Their spacetime transformations are the standard ones for conformal primaries under inversion and reflection,
 \beq
 \begin{aligned}
&\phi(x)\mapsto \phi(x')\,,\quad \psi(x) \mapsto i \frac{\gamma_{\mu} x^{\mu}}{|x|} \psi(x')\,,\\
&A_{\mu} (x) \mapsto - I_{\mu}{}^{\nu} A_{\nu}(x')\quad {\rm{with}}\quad I_{\mu}{}^{\nu}= \delta_{\mu}^{\nu} - 2\frac{x_{\mu} x^{\nu}}{x^2}\,.
\end{aligned}
\eeq
Additionally, their R-symmetry transformation induced by (\ref{halfBPS}) is dictated by consistency of the superconformal transformation, namely
\beq
\delta_{\epsilon} A_{M} = \epsilon \Gamma_{M} \Psi\, ,
\eeq
with the transformation of the Killing spinor given in (\ref{killingtransform}).
We will write the explicit transformations of the elementary fields below.

Finally, we may also
transform the gauge group generators under the antipodal map. 
The SU($N$) gauge group is  endowed with an outer automorphism $\tau$ that acts as complex conjugation of an element $g$ of SU($N$), $\tau: g\mapsto g^{*}$. Clearly $\tau^2=1$ and this involution customarily referred to as \textit{charge conjugation}. In this paper, we are using anti-hermitian generators $T_a$ for SU($N$) (see appendix \ref{conventionslag} for our conventions), so that
\beq \label{outer}
\tau: (T_a)^{m}{}_{n} \mapsto -(T_{a})^{n}{}_{m}\,.
\eeq

We conclude that we have two distinct choices for specifying $\mathcal{N}=4$ SYM on $\mathbb{RP}^{4}$ depending on whether we combine or not the spacetime and R-symmetry transformations with the charge conjugation $\tau$.

\paragraph{Without charge conjugation.} The elementary $\mathcal{N}=4$ SYM fields are identified as
\beq \label{identi}
\Phi_{I}^{a} (x')T_{a}= \hat\Phi_{I}^{a}(x)T_{a}\,,\quad A^{a}_{\mu} (x')T_{a}= - I_{\mu}{}^{\nu} A^{a}_{\nu}(x)T_{a}\,,\quad \Psi^{a}(x')T_{a}=-i \frac{\tilde\Gamma_{\hat\mu}x^{\mu}}{|x|} \mathcal{R} \Psi^{a}(x)T_{a}\,.
\eeq
where $\hat \Phi_{I}=\left( \Phi_{5},\Phi_{6},-\Phi_{7},\Phi_{8},-\Phi_{9},-\Phi_{0}\right)$.
A new feature of $\mathcal{N}=4$ SYM on $\mathbb{RP}^{4}$ is that conformal symmetry no longer prevents some operators to develop vevs. In particular, spinless operators which are even under the involution $\mathcal{I}_{\text{SYM}}$ might have a nontrivial one point function whereas all other operator remain vevless as a result of the selection rules that follow from (\ref{symm}).
In perturbative field theory, one can explicitly see these vevs arising from the interaction of a field with its image given by (\ref{identi}). For example, a scalar propagator in $\mathbb{RP}^{4}$ is given by
\beq
\langle[\Phi_{I}]^{m}{}_{n} (x) [\Phi_{J}]^{p}{}_{q}(y)\rangle = \frac{g^{2}_{\rm{YM}} \delta_{IJ}}{16 \pi^2} \left(\frac{1}{\eta}\pm \frac{1}{1-\eta} \right) \left( \delta^{m}_{q} \delta^{p}_{n} -\frac{1}{N} \delta^{m}_{n}\delta^{p}_{q}\right)
\eeq
where $m,n,\dots$ are gauge indices, $\eta = (x-y)^2/(1+x^2)(1+y^2)$ is the chordal distance between two points on $S^4$ and the sign follows from the parity of the scalar field under the involution as defined below the equation (\ref{identi}). We provide more details on the $\mathbb{RP}^4$ propagators in the appendix \ref{appprops}. The second term is absent in the usual $S^4$ but it is present here and gives a finite contribution as $y\rightarrow x$.

\begin{figure}[t]
\centering
  \includegraphics[scale=.31]{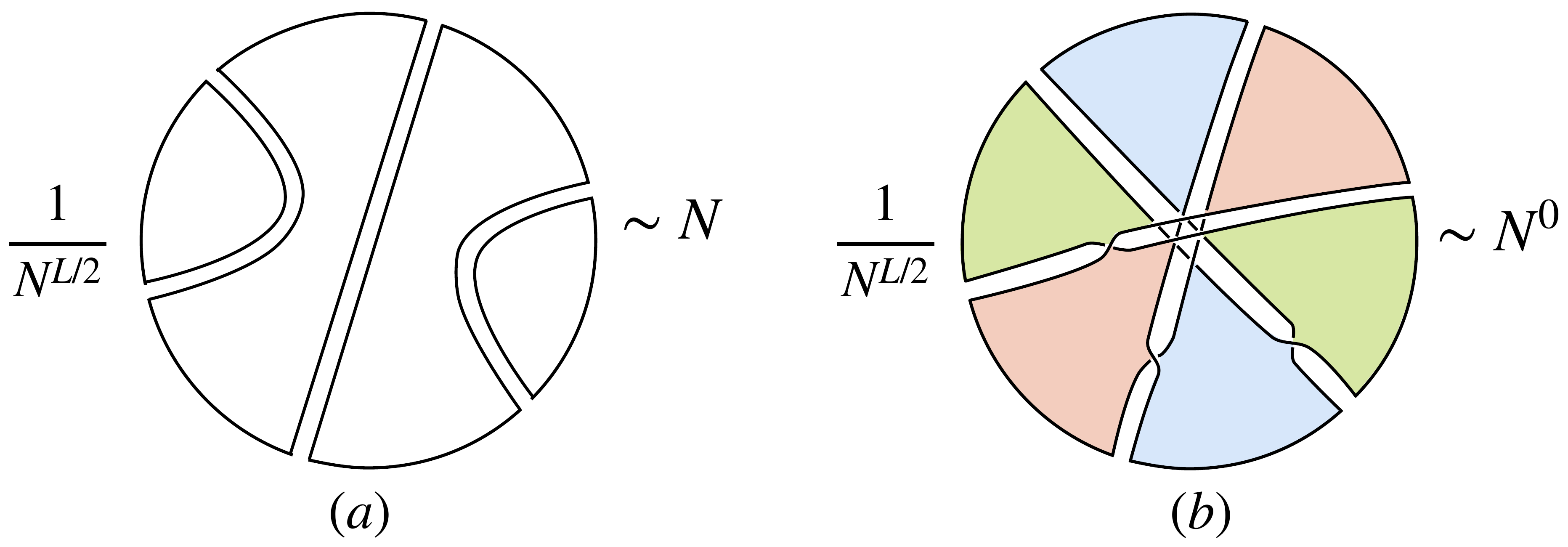}
  \caption{$(a)$ A leading diagram in double line notation contributing to the one-point function of a single trace operator of length $L=6$ in the large $N$ limit of $\mathcal{N}=4$ SYM on $\mathbb{RP}^4$ without the gauging of charge conjugation. The pre-factor ensures the UV limit of the two-point functions is normalized to one. $(b)$ A leading diagram in the large $N$ limit contributing to the one-point function of the same single trace operator in $\mathcal{N}=4$ SYM on $\mathbb{RP}^4$ with the gauging of charge conjugation. In this example, the double line notation graph originates a surface with three faces (illustrated in distinct colors) as opposed to the four faces of the figure (a), hence producing a different large $N$ scaling.}
  \label{fig:feyng}
\end{figure}

Let us discuss the large $N$ scaling of the vevs.  We consider a single-trace operator which we write schematically as
\beq
\mathcal{O} \sim \Tr [\chi_{1} \dots \chi_{L}] 
\eeq
for generic fields $\chi_1, \dots, \chi_{L}$ such that $\mathcal{O}$ is a Lorentz scalar and also singlet under SO(3)$\times$SO(3). The one point functions can be computed in field theory using the scalar propagators above together with the fermion and gauge field propagators from the appendix \ref{appprops}. A typical Feynman diagram contributing to the one point function at the leading order in the large $N$ 't Hooft expansion is represented in figure \ref{fig:feyng}$(a)$. When the two point functions of $\mathcal{O}$ and its conjugate  are normalized to the unit norm in the UV limit, then the corresponding vev has the following large $N$ scaling
\beq \label{nochargescale}
\langle \mathcal{O} \rangle \sim O(N)\,.
\eeq We will provide explicit examples of this computation at tree level in the section \ref{sectree}. This the standard scaling expected from a classical gravitational background and the goal of this paper is to determine a new solution of type IIB supergravity dual to this setup.

\paragraph{With charge conjugation.} If we accompany the involution $\mathcal{I}_{\rm{SYM}}$ with the charge conjugation\footnote{We thank Shota Komatsu for many ideas and discussions leading to this setup.} $\tau$, the transformation of elementary fields is now given by
\beq \label{identi}
\Phi_{I}^{a} (x')T_{a}= -\hat\Phi_{I}^{a}(x)T^{\top}_{a}\,,\quad A^{a}_{\mu} (x')T_{a}=  I_{\mu}{}^{\nu} A^{a}_{\nu}(x)T^{\top}_{a}\,,\quad \Psi^{a}(x')T_{a}=i \frac{\tilde\Gamma_{\hat\mu}x^{\mu}}{|x|} \mathcal{R} \Psi^{a}(x)T^{\top}_{a}\,.
\eeq
The transposition of the SU($N$) generators changes the large $N$ scaling of the one-point functions. As a consequence, the scalar propagator now reads
\beq
\langle[\Phi_{I}]^{m}{}_{n} (x) [\Phi_{J}]^{p}{}_{q}(y)\rangle = \frac{g^{2}_{\rm{YM}} \delta_{IJ}}{16 \pi^2} \left(\frac{\left( \delta^{m}_{q} \delta^{p}_{n} -\frac{1}{N} \delta^{m}_{n}\delta^{p}_{q}\right)}{\eta}\mp \frac{\left( \delta^{mp} \delta_{n q} -\frac{1}{N} \delta^{m}_{n}\delta^{p}_{q}\right)}{1-\eta} \right)\,,
\eeq
and similarly for the remaining elementary fields.
Ignoring the $1/N$ part of the expression above, the color indices have the same structure as for the SO($N$) gauge group. As well known, the  SO($N$) gauge group produces a large $N$ 't Hooft  expansion in powers of $1/N$ rather than $1/N^2$. In contrast to the previous case, a Feynman diagram contributing to the one-point function at leading order in the large $N$ limit is shown in figure \ref{fig:feyng}$(b)$. In this example, the fields at the positions $n$ and $n +L/2$ inside the trace  are connected by a propagator and this maximizes the number of faces of the corresponding surface obtained by the double line notation diagram. As a consequence, we have that the leading term contributing to the single trace vev is rather
\beq
\langle \mathcal{O} \rangle \sim O(1)\,.
\eeq 
This is strikingly distinct from (\ref{nochargescale}). This result is suggestive and we can anticipate some consequences for the holographic dual of this configuration.
To build up some intuition, we can start by extending the boundary involution into the bulk and use it to construct a $\mathbb{Z}_2$ orbifold of $EAdS_{5}\times dS_5$. In embedding coordinates, $EAdS_5$ can be parametrized by
\beq
X^{0}=L \cosh r\,,\quad X^{i}=L \sinh r\, \Omega^{i}\quad i=1,\dots,5\,,
\eeq
with $\Omega^i$ being the a unit vector embedding $S^4$. In these coordinates, the Euclidean AdS metric is expressed as
\beq
ds^2 = L^2 \left(dr^2+\sinh^2 r \, ds^2_{S^4} \right)\,.
\eeq 
The five-dimensional de Sitter is parametrized by the coordinates $Y^{I=5\dots 9,0}$ satisfying $\sum_{I=5}^{9}\left(Y^{I}\right)^2-(Y^{0})^2 = 1$ with the metric $ds^2_{dS_5}=L^2\eta_{IJ}\,dY^{I}dY^{J}$ with $\eta =\text{diag}(+++++-)$.
We extend the boundary identification (\ref{invol1}) to $EAdS_{5} \times dS_5$  in analogous manner 
\beq \label{adsinv}
X^{0} \sim X^{0}\quad X^{i}\sim -X^{i}
\eeq
together with 
\beq
Y^{I} \sim \hat{Y}^{I}
\eeq
as induced by the transformation of SYM scalars given in (\ref{identi}). This new bulk involution 
 \beq \label{bulkinv}
 \mathcal{I}:(X^{0},X^{i},Y^{I})\mapsto (X^{0},-X^{i},\hat{Y}^{I})
 \eeq has a fixed locus at
 \beq \label{locus}
 r=0\,,\quad Y^{5}=Y^{6}=Y^{8}=0\,,\quad (Y^{7})^2+(Y^{9})^2-(Y^{0})^2=1\,.
 \eeq
 A salient feature of $\mathcal{I}$ is that the volume forms of both $EAdS_5$ and $dS^5$ are odd under such map. This would then force the self-dual five-form flux 
 \beq 
 F_{5} =\frac{4}{L}\left(d\text{vol}_{EAdS_5}+ d\text{vol}_{dS_5} \right)
 \eeq  to vanish unless we combine it with an orientation-reversal on the worldsheet, often called parity and denoted by $\Omega$. 
In fact, $\Omega$ is precisely implemented  on the gauge theory side by the outer automorphism $\tau$ of the gauge group SU($N$). The resulting picture is that of an  orientifold at the fixed-locus of the involution $\mathcal{I}$, namely a O1 plane spanning the $dS_2$ parametrized by $(Y^{7})^2+(Y^{9})^2-(Y^{0})^2=1$. 
The  orientifold introduces a crosscap on the worldsheet, which modifies the  string topological expansion and results in the above scaling of the one-point functions. We will study  this setup in a forthcoming publication \cite{inprogresscc}. In the rest of the paper we will focus on the holographic dual of the setup {\it not} involving the charge conjugation of the gauge group.

\section{Holography} \label{holo}

The objective of this paper is the construction of 
 dual  background to the setup ``without charge conjugation'' described above. Before proceeding with our main task, let 
us comment briefly on the dual description of the setup ``with charge conjugation''. In that case, the $O(1)$ scaling of one-point functions
means that the dual classical  geometry is  unchanged, up to  the need to extend into the bulk the involution defining $\mathbb{RP}^4$. This amounts 
to a $\mathbb{Z}_2$ orbifold of $EAdS_{5}\times dS_{5}$, which contains a fixed locus of dimension two specified by (\ref{locus}). Therefore, a bulk spacetime singularity  is expected at the center of $AdS$ ($r=0$) and $\theta= \pi/2$ (which is the fixed point of the map $\theta \mapsto \pi-\theta$) at which point $S^2$ shrinks to zero.
The involution $\tau$, which maps the fundamental representation to its complex conjugate, is the gauge theory counterpart of the worldsheet parity $\Omega$ that reverses the string orientation. Therefore, the orbifold should be complemented with an additional gauging of $\Omega$ leading to the picture of an  orientifold O$1$ plane. As a check, the orientifold 
is in fact needed in order for 
the RR four-form (whose five-form flux supports the background) to survive the involution.
 In summary, for the setup with charge conjugation there is a compelling candidate for the dual description, in terms of an orientifold of the standard $AdS_5 \times S^5$ background.

By contrast, in the setup ``without charge conjugation'' the large $N$ scaling of one-point functions 
tells us that
we need to look for a  completely different classical solution. To this end, we
 will employ a by now standard strategy of considering a consistent truncation of type IIB on $S^5$ provided by the five dimensional $\mathcal{N}=8$ gauged supergravity \cite{Gunaydin:1985cu}. Within this lower dimensional supergravity theory, we  will construct a solution compatible with the symmetries expected from the corresponding field theory and then uplift back to ten dimensions. Since we are interested in a Euclidean solution, it will be crucial to perform an analytic continuation from the standard Lorentzian formulation of supergravity. 

\subsection{5D Gauged Supergravity Truncation}
In Lorentzian signature, $\mathcal{N}=8$ gauged supergravity contains a gauge group SO(6) that is identified under the AdS/CFT correspondence to the SO(6)$_{R}$ of the gauge theory. Therefore, the R-symmetry breaking pattern of the dual field theory like the one we are interested (\ref{symm}) (although in Euclidean signature) dictates the field content that should be kept in supergravity to respect the desired symmetries. 

Besides the metric, the remaining bosonic fields of this supergravity theory are organized by the SO(6)$\times$SO(2) subgroup of the global symmetry E$_{6(6)}$. It comprises the fifteen SO(6) gauge fields (after gauging SO(6) $\subset$ E$_{6(6)}$), twelve two-forms transforming in the fundamental of SO(6)$\times$SO(2) and forty two scalars that transform according to
\beq \label{reps}
{\bf{20}}'_{(0)} \oplus {\bf{10}}_{(-2)}  \oplus \bar{\bf{10}}_{(2)} \oplus {\bf{1}}_{(4)}  \oplus {\bf{1}}_{(-4)} \,.
\eeq
Each of these representations can be identified with specific operators in $\mathcal{N}=4$ SYM by matching the corresponding R-symmetry representations.
As  mentioned above, we select among this vast field content only the particular fields that are singlets under the subgroup of SO(6) of interest. 
In the present context, we have in addition to account for the fact that we will be interested in a solution with Euclidean signature which means that the R-symmery group is SO(5,1) rather than SO(6). In practice, that can be easily implemented by complexifying a particular  scalar as we will explain below.

Let us first describe the bosonic subsector compatible with SO(3)$\times$SO(3) R-symmetry, which we will then analytically continue to the one of interest (\ref{symm}). The embedding of this subgroup into SO(6) is specified by the following decomposition of the fundamental representation ${\bf{6}} \rightarrow  ({\bf 3},{\bf 1})+ ({\bf 1},{\bf 3}) $.  The above representations (\ref{reps}) branch according to
\beq
\begin{aligned}
{\bf{20}}'& \rightarrow ({\bf 1},{\bf 1})+({\bf 3},{\bf 3})+({\bf 5},{\bf 1})+({\bf 1},{\bf 5}) \\
{\bf{10}},{\bf{\bar{10}}}& \rightarrow ({\bf 1},{\bf 1})+({\bf 3},{\bf 3})  \\
{\bf{1}}& \rightarrow ({\bf 1},{\bf 1})\,.
\end{aligned}
\eeq
We keep the five singlets resulting from this branching and name them as follows
\begin{alignat}{2} \label{trunc5}
&\alpha:\quad &&{\bf{20}}'_{0} \rightarrow ({\bf 1},{\bf 1})_{0}\\
&\chi\, e^{-i \omega}: \quad &&{\bf{10}}_{-2} \rightarrow ({\bf 1},{\bf 1})_{-2}\\
&\chi \, e^{i \omega}: \quad &&\bar{{\bf{10}}}_{2} \rightarrow ({\bf 1},{\bf 1})_{2}\\
&\varphi \, e^{ i c}:\quad &&{\bf{1}}_{4} \rightarrow ({\bf 1},{\bf 1})_{4}\\
&\varphi \,e^{- i c}:\quad  &&{\bf{1}}_{-4} \rightarrow ({\bf 1},{\bf 1})_{-4}\,.
\end{alignat}
We can identify $\varphi$ and $c$ as the five-dimensional dilaton and axion dual of the Yang-Mills coupling and $\theta$ angle respectively and
\beq \label{dualops}
\begin{aligned}
\alpha &\leftrightarrow   \sum_{I=1}^{3} \tr  \,  (\Phi^{I})^2  -\sum_{I=4}^{6} \tr \, (\Phi^{I})^2 \\
\chi \, e^{ i \omega} &\leftrightarrow \tr \Big[ c_1\, C_{123}^{A B}\lambda_{A}\lambda_{B}+c_2\, C_{456}^{A B}\lambda_{A}\lambda_{B}+c_{3} X_{1} [X_{2}, X_{3}]+c_{4} X_{4}[ X_{5}, X_{6}] \Big]
\end{aligned}
\eeq
where in the last line $c_{1,2,3,4}$ are (generally complex) coefficients that can be fixed by diagonalizing the mixing matrix but we will not need them in this work.
In addition to these scalars, only the metric will be turned on and all gauge fields, two-forms and remaining scalars are consistently set to zero. 
It turns out that this truncation to a five scalar model coincides with the one used to construct the holographic dual of the superconformal Janus interface in $\mathcal{N}=4$ SYM with $\mathcal{N}=4$ supersymmetry \cite{Bobev:2020fon}. In particular, there are eight generators of E$_{6(6)}$ commuting with residual SO(3)$\times$SO(3) which close into an $\mathfrak{sl}(3,\mathbb{R})$ algebra. The five scalars above are then the coordinates in the coset
\beq \label{coset}
\text{SL}(3,\mathbb{R})/\text{SO}(3)\,.
\eeq
We will closely follow \cite{Bobev:2020fon} and choose the representative 27-bein in E$_{6(6)}$ parametrizing this coset in a similar fashion (we refer the reader to the appendix \ref{appuplift} and appendix A of \cite{Bobev:2020fon} for further details). 

\subsection{Lagrangian and BPS equations}
We choose the metric and scalars to be compatible with the SO(5) symmetry of $\mathbb{RP}^4$. Given that $\mathbb{RP}^4$ is a quotient of $S^4$ by an isometric involution, it inherits its metric as well. We use the following metric ansatz
\beq \label{metric}
ds^2_{\text{5D}}= dr^2 +e^{2 A} ds^2_{\mathbb{RP}^4}
\eeq
where for $ds^2_{\mathbb{RP}^4}$ we take the round metric of the unit $S^4$. We assume that the warp factor $A$ together with the five scalars are functions of the radial coordinate $r$ only,
\beq
A=A(r)\,,\quad \alpha=\alpha(r)\,,\quad \chi=\chi(r)\,,\quad \varphi=\varphi(r)\,,\quad \omega=\omega(r)\,,\quad c=c(r)\,.
\eeq
It is now a simple matter to derive the Lagrangian and BPS equations using for example the parametrization of the coset  (\ref{paramU}) detailed in  appendix \ref{appuplift}, by implementing the formulae in \cite{Gunaydin:1985cu}. Since we are after an Euclidean solution, we have in addition  to account for the fact that R-symmetry has to be continued to SO(5,1). The strategy to construct Euclidean solutions from a Lorentzian supergravity theory has been considered in various examples \cite{Freedman:2013oja,Bobev:2013cja,Bobev:2016nua,Bobev:2018ugk,Bobev:2020pjk}. In our present context, it amounts to analytically continue a scalar and also the time component of the 5D gamma matrices (since we are in a mostly plus signature)
\beq \label{ancont}
\chi \rightarrow i \chi\,, \quad \quad \gamma_{0} \rightarrow -i\gamma_0\,.
\eeq
In the Lagrangian, the effect of this continuation is minimal. The five dimensional Lagrangian can be recycled from \cite{Bobev:2020fon} and after implementing (\ref{ancont}), it is given by
\beq \label{fulllag}
\mathcal{L}_{\text{5D}} = \frac{1}{16 \pi G_{5}} \left(R_{5}+ \mathcal{K}  - V(\alpha,\chi) \right)
\eeq
where $R_{5}$ is the five-dimensional Ricci scalar, $\mathcal{K}$ is the kinetic term, and $V(\alpha,\chi)$ is the scalar potential that depends only on two scalars and reads
\beq
V(\alpha,\chi) = -\frac{3 g^2}{4}  (\cosh 4 \alpha \, \cos 4 \chi +3)\,,
\eeq
where $g$ is the gauge coupling.
The kinetic term $\mathcal{K}$ depends very non-linearly on all scalars and its full expression will be shown below after explaining some simplifying features.

In principle, all five scalars might acquire a non-trivial profile in the bulk. We expect that $\chi, \alpha$ decay asymptotically for large $r$ and the remaining scalars become constants, so that in the UV we obtain an asymptotically AdS solution. This is expected as the short distance behaviour of the boundary gauge theory is  unaffected by the involution $\mathcal{I}_{\text{SYM}}$.
However, there are further constraints to be imposed from the fact that the  gauge theory is specifically living on $\mathbb{RP}^4$.
For example, the $\theta$ angle of Yang-Mills defined on a unorientable manifold such as $\mathbb{RP}^4$ is limited to take up either the value $0$ or $\pi$. In this paper, we will focus on the case of $\theta =0$. The value of the  $\theta$ angle is dual to the ten-dimensional axion which in turn is related to the five dimensional axion $c(r)$. The precise relation is determined in the appendix \ref{appuplift} from the ten-dimensional uplift formulae. Therefore, we expect  a fixed value for the asymptotic value of $c(r)$ as $r\rightarrow \infty$ (the location of the boundary) consistent with the boundary  $\theta$ angle.

Another related constraint can be obtained from the expectation value of the operators in the representation $\bf{10}+\bf{\bar{10}}$ dual to the scalars $\chi e^{\pm i \omega}$. It turns out that the phase of the expectation value of such operators is fixed by CPT invariance of the gauge theory. That again sets the value of the boundary behaviour of $\omega$ to a certain constant. In  appendix \ref{sugradets}, starting from the most general supergravity equations we show that we can consistently set both $c(r)$ and $\omega(r)$ scalars to constants in the bulk: the five-dimensional axion takes the value $c(r)=0,\pi$ but both these choices lead to the same ten-dimensional solution and hence are physically equivalent; for $\omega(r)$, we have a discrete choice to make $\omega(r)=0$ or $\pi/2$. The two choices lead to two distinct solutions related by S-duality. For most of the main text, we will discuss in detail the choice $\omega(r)=\pi/2$ and in the section \ref{secsdual} we will comment on the S-dual solution corresponding to $\omega(r)=0$.

With these simplifications, the kinetic term $\mathcal{K}$  of the five-dimensional Lagrangian gets reduced to
\beq
\mathcal{K} =8 \chi '^2-\frac{3}{4} \alpha '^2 (3 \cos (8 \chi )+5)-\frac{1}{4} \varphi '^2 (\cos (8 \chi )+7)
-3\, \alpha '  \varphi '  \sin ^2(4 \chi )\,,
\eeq
where  prime denotes derivatives with respect to $r$ and the \textit{wrong} sign of the kinetic term for the scalar $\chi$ results from the analytic continuation (\ref{ancont}).

Let us now discuss the BPS equations with these two scalars set to their boundary values for any $r$, $\omega(r)=\pi/2$ and $c(r)=
\pi$. Following appendix \ref{sugradets}, the spin-1/2 BPS equations become
\beq \label{spin12}
\begin{aligned}
i( \alpha'+\varphi')\, \gamma_{r} \,\epsilon^{s} &= \frac{1+\sqrt{2}}{6} \partial_{\alpha}\widetilde{\mathcal{W}}\, \epsilon^{\bar{s}}\\
( \alpha'+\varphi')^2&= \frac{1}{36} \partial_{\alpha} \mathcal{W}\, \partial_{\alpha} \widetilde{\mathcal{W}}\\
\chi' ( \alpha'+\varphi')&= \frac{1}{24} \sin 4\chi \left( \frac{\mathcal{W}\, \partial_{\alpha} \mathcal{W} +\widetilde{\mathcal{W}}\, \partial_{\alpha}\widetilde{\mathcal{W}}}{2} \right) \\
\varphi'( \alpha'+\varphi')&=-\frac{1}{24} \tan 4\chi \left( \frac{\mathcal{W} \, \partial_{\alpha} \mathcal{W}-\widetilde{\mathcal{W}}\, \partial_{\alpha}\widetilde{\mathcal{W}}}{2}\right)
\end{aligned}
\eeq
with $\mathcal{W},\widetilde{\mathcal{W}}$ being the real superpotentials given in (\ref{superpots}), $\gamma_{r}$ is the five-dimensional gamma matrix along the $r$ direction and $\epsilon^{s,\bar{s}=1,2,3,4}$ are symplectic-Majorana spinors parametrizing the sixteen independent real supercharges.

The last three equations can be used to show that the quantity
\beq \label{const1}
\mathcal{J} \equiv 2\frac{\left(\cosh 4\alpha \cos 4\chi -1\right)}{\sin^{4/3} 4\chi}
\eeq
is conserved, $ \mathcal{J}'=0$. This constant as we will see controls the asymptotic value of both $\alpha$ and $\chi$ scalars towards the boundary and therefore it is directly related to the strong coupling limit of the one-point function of the corresponding dual operators.

In order to make further progress in solving these equations with unconstrained spinors $\epsilon^{s,\bar{s}}$ so that we preserve sixteen supercharges, we combine them with the spin-3/2  BPS equations determined in appendix \ref{sugradets}. 
Using the first equation in (\ref{spin12}) together with (\ref{spin32}), we arrive at
\beq \label{warpeqs}
 e^{-A}+A' =-\frac{\mathcal{W}\, \partial_{\alpha} \widetilde{\mathcal{W}}}{18(\alpha'+\varphi')}\,, \quad (A')^2 = \frac{\mathcal{W} \widetilde{\mathcal{W}}}{9}+e^{-2A}\,
\eeq
The two sets of equations (\ref{spin12}) and (\ref{warpeqs}) share a few similarities with the corresponding equations determining the supergravity background  dual to the $\mathcal{N}=4$ Janus interface \cite{Bobev:2020fon}. However, we will see that our solution will be sharply distinct and with a completely different physical interpretation.

\subsection{Solution of BPS equations}
\begin{figure}[t]
\centering
  \includegraphics[scale=.25]{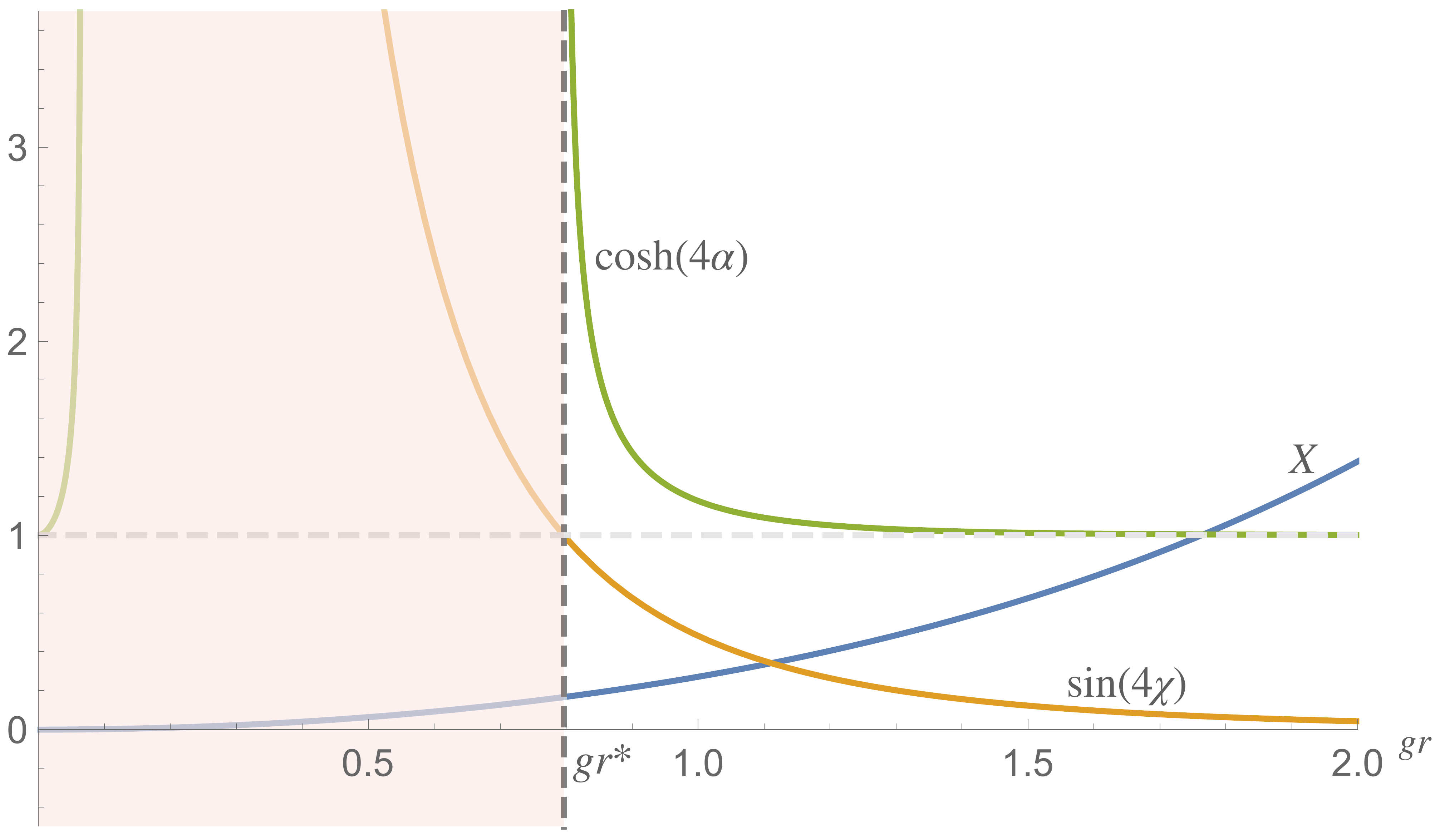}
  \caption{This plot exhibits the profile of the scalars $\chi$ and $\alpha$ and the warp factor $X$ as function of the radial coordinate $r$ for $\mathcal{J}=1/6$. Importantly, the region for $r\leq r^{\ast}$ depicted in red is not physical: the scalar $\alpha$ develops a singularity  precisely at $r=r^{\ast}$ and $\chi$ becomes complex for $r\leq r^{\ast}$.}
  \label{fig:scalars}
\end{figure}
Using the equations (\ref{warpeqs}) and the constant (\ref{const1})  we can express the scalars in terms of the warp factor
\beq
\begin{aligned} \label{scals}
\sin4 \chi  &= \,\left( \frac{\mathcal{J}}{X} \right)^{3/2}\\
\cosh 4 \alpha&= \frac{2 X^2+\mathcal{J}^3}{2\sqrt{X\left(X^{3}- \mathcal{J}^3 
\right)}} \, ,
\end{aligned}
\eeq
where $X = \frac{g^2}{4}\, e^{2A}$ and we take $\mathcal{J} \geq 0$ to ensure real scalars. As we will see in the next section, complex scalars would render the uplifted ten-dimensional metric complex. Finally, we use the second equation in (\ref{warpeqs}) and (\ref{scals}) to determine the warp factor
\beq
\frac{4}{g^2} (X')^2= \mathcal{J}^3 + 4 X (X+1) \,.
\eeq
This equation is quadratic and admits two solutions related by $r\rightarrow -r$. We pick the solution
\beq
X(r) = \frac{\mathcal{J}^{3}}{4}  \sinh g r+\frac{1}{4} (2-\mathcal{J}^{3}) \cosh g r-\frac{1}{2}\,,
\eeq
where we have adjusted the origin of $r$ so that the solution asymptotes to $AdS_5$ with radius $2/g$. This solution completely determines the bulk profile of the scalars $\chi$ and $\alpha$. It is apparent from (\ref{scals}) that they become singular or complex for $r\leq r^{\ast}$ where $r^{\ast}$ is defined by
\beq \label{rstar}
 X(r^{\ast}) =\mathcal{J}\,,
 \eeq 
 see the plot in figure \ref{fig:scalars}. In fact, that region is ill defined and thus unphysical and this is the first sign that the solution contains a bulk singularity.

\begin{figure}[t]
\centering
  \includegraphics[scale=.25]{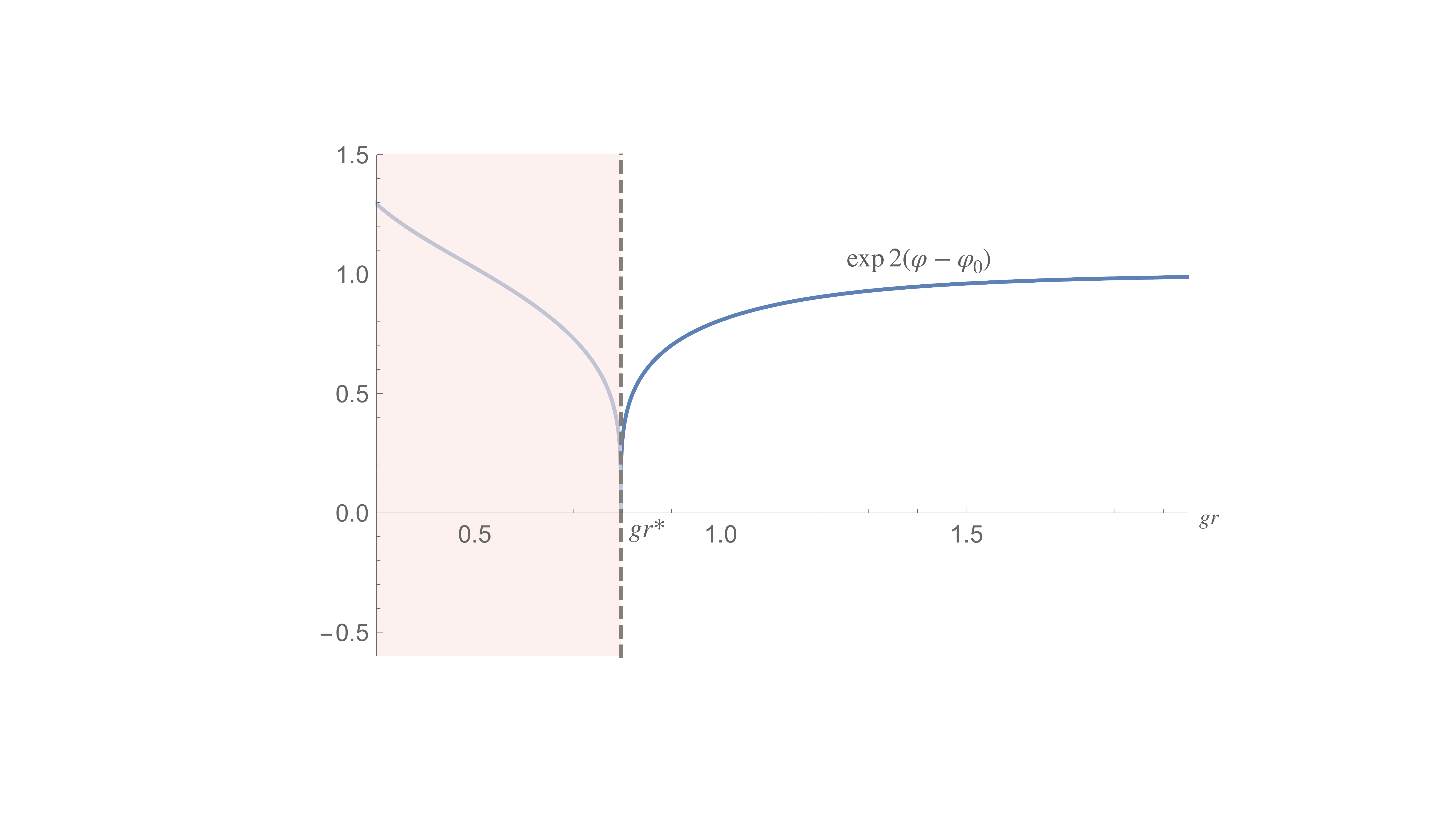}
  \caption{This plot exhibits the profile of the five-dimensional dilaton $\varphi$ as function of the radial coordinate $r$ for $\mathcal{J}=1/6$. At $r= r^{\ast}$, the five-dimensional dilaton becomes singular $\varphi \rightarrow -\infty$. The region depicted in red is not physical.}
  \label{fig:scalars2}
\end{figure}

The remaining non-trivial scalar $\varphi$ is determined using (\ref{spin12}) and we get
\beq
\exp{ (\varphi -\varphi_{0})} = \exp{\left[ \int_{\infty}^{X} dx \frac{3\, \mathcal{J}^{3/2}(\mathcal{J}^{3} + 2 x^2) }{ 8x \left(x^3-\mathcal{J}^{3} \right) \sqrt{\mathcal{J}^{3} +4 x (x+1)}}\right] }\,,
\eeq
where  $\varphi_{0}$ is the asymptotic value of the 5D dilaton for $r\rightarrow \infty$. The integral can be computed explicitly and we show the dilaton profile in figure \ref{fig:scalars2}. At the point $r=r^{\ast}$, the five-dimensional dilaton becomes singular ($\exp \varphi \rightarrow 0$), indicating again that the solution for $r\leq r^{\ast}$ becomes unphysical. 

\subsection{Boundary behaviour}
To extract the UV asymptotics, we rewrite the metric in the Fefferman-Graham form
\beq \label{FG}
ds^2_{\text{5D}}= \frac{4}{g^2} \left[ \frac{d\rho^2}{4 \rho^2} +\frac{1}{\rho}\left( \frac{1}{4}-\frac{\rho}{2} +\frac{\rho^2}{4}(1-\mathcal{J}^3) \right)ds^2_{\mathbb{RP}^4} \right]
\eeq
where $\rho \equiv e^{-gr}$ with the boundary being located at $\rho = 0$.
The scalars admit the following expansions
\beq
\begin{aligned}
\alpha &\simeq \mathcal{J}^{3/2} \rho +4 \,\mathcal{J}^{3/2}\rho^2   
+\mathcal{O}(\rho^3) \\
\chi &\simeq 2\, \mathcal{J}^{3/2} \rho ^{3/2}+6\, \mathcal{J}^{3/2} \rho ^{5/2}+\mathcal{O}(\rho^{7/2})\\
\varphi &\simeq \varphi_0 -3\, \mathcal{J}^{3/2}\rho^2+\mathcal{O}(\rho^4)\,.
\end{aligned}
\eeq
 
We find that our solution is characterized by three parameters $g, \varphi_{0}$ and $\mathcal{J}$. The supergravity gauge coupling $g$ is identified with the AdS radius $L$ by the relation $1/g= \frac{L}{2}$. We read off from the metric (\ref{FG})
that $L/2$ is also the radius of the boundary metric. 
In the standard setup where the boundary is conformally flat (e.g.~$S^4$) and vevs vanish, the dimensionful parameter $L$ would drop out of all physical observables, but this is of course not the case for $\mathbb{RP}^4$, where $L$ sets the scale of the one-point functions.
By the standard AdS/CFT dictionary (see e.g.~\cite{Skenderis:2002wp}), the expected form of the asymptotic expansion of a generic scalar $\phi$ is as follows, 
\beq
\begin{aligned} \label{bdryexp}
\phi(\rho) &= \rho^{\frac{4-\Delta}{2}} ( \phi_{s}+\dots)+ \rho^{\frac{\Delta}{2}} ( \phi_{v}+\dots)\quad \text{for }\,\Delta \neq 2\,,\\
\phi(\rho) &= \rho \left( \log\rho\, (\phi_{s}+\dots)+   \phi_{v}+\dots\right)\quad \text{for }\,\Delta = 2\, ,
\end{aligned}
\eeq
where $\phi_{s}$ is related to a deformation of the Lagrangian  by the dual boundary operator  and  $\phi_{c}$ to its vev, and the dots correspond to higher order corrections in the radial distance.
 The constant $\varphi_{0}$ controls the asymptotic value of the 5D dilaton and therefore is related to the Yang-Mills coupling $g_{\text{YM}}$. The precise relation can be obtained from the ten dimensional uplifted dilaton that will be worked out in the next section. The expectation from the field theory side is
that once we fix the radius of $\mathbb{RP}^4$, the rank of the gauge group (controlled by the RR flux), the Yang-Mills coupling and the theta angle (which we have set to zero), there should be no other free parameters -- i.e., one-point functions should be completely determined by the dynamics.  Instead, we find a continuous family of solutions
parametrized by the constant $\mathcal{J}$, which shows up as the leading term in the asymptotics of the scalars $\alpha$ and $\chi$. Since there are no source terms for the Lagrangian deformation in these expansions, 
 this constant controls the vevs of the corresponding field theory operators. In the next section
we will comment on how this last parameter could in principle be fixed.

\section{Uplift to type IIB supergravity} \label{uplift}
In this section, we present the uplift of the previous solution to ten dimensional type IIB supergravity. This is done by making extensive use of the formulae developed in \cite{Baguet:2015sma} adapted to our Euclidean case. Some details of this uplift are provided in the appendix \ref{appuplift}. Here, we simply present the ten dimensional solution.
In order to realize the residual R-symmetry SO(3)$\times$SO(2,1) of the boundary theory, 
we will make use of $S^2$ and $dS_2$ slices with the metric
\beq
d\Omega^{2}_{S^2} = d\phi_1^2+\sin^{2}\phi_1 d\phi_2^2\,\quad\quad d\Omega^{2}_{dS_2} =-d\chi_1^2+\cosh^{2}\chi_1 d\chi_2^2\,.
\eeq
We start with the ten-dimensional metric. In the Einstein frame, it is given by
\beq \label{tendmetric}
ds^2_{\text{10D}}=\Delta^{1/4}\left(ds^2_{\text{5D}}+\frac{4}{g^2}\left(d\theta^2+ \frac{\cos^2 \theta}{1+\mathcal{K}_{+}\cos^{2}\theta}d\Omega^{2}_{S^2} + \frac{\sin^2 \theta}{1+\mathcal{K}_{-}\sin^{2}\theta }d\Omega^{2}_{dS_2}  \right) \right)
\eeq
where $\mathcal{K}_{\pm}= e^{\pm4 \alpha} \cos 4 \chi-1$ and the prefactor $\Delta$ is given by
\beq
\Delta = \left(\mathcal{K}_{-} \sin ^2(\theta )+1\right) \left(\mathcal{K}_{+} \cos ^2(\theta )+1\right)\,.
\eeq
The solution also contains both NSNS and RR two-form potentials. They have the following expressions,
\beq
B_2=-\frac{4 i}{g^2}\frac{e^{\alpha +\varphi}\cos \theta \sin 4\chi}{1+ \mathcal{K}_{+}+\tan^2 \theta} dV_{S^2}\,, \quad\quad C_{2} = \frac{4}{g^2} \frac{e^{-\alpha -\varphi} \sin \theta \sin 4\chi}{1+\mathcal{K}_{-}+\cot^2\theta}dV_{dS_2}
\eeq
where $dV_{S^2}$ and $dV_{dS_2}$ are the volume forms on $S^2$ and $dS_2$ respectively
\beq
dV_{S^2}= \sin \phi_1 d\phi_1 \wedge d\phi_2\,, \quad \quad dV_{dS_2}= \cosh \chi_1 d\chi_1 \wedge d\chi_2\,.
\eeq
The RR four-form potential reads\footnote{We remark that the uplift formulae provides only a part of the five-form in type IIB, namely $\mathcal{F}_{5}= dC_{4}-\frac{1}{2}\left(C_{2}\wedge dB_2 -B_{2}\wedge dC_2 \right)$. To make the five-form self dual we have to explicitly add  $\star \mathcal{F}_5$, so that the full result is $F_5=\mathcal{F}_5+\star\mathcal{F}_5$.}
\beq
C_{4} = \frac{2i}{g^4} \left(  \frac{\sin^{3}2\theta}{\Delta} \left( \mathcal{K}_{+}\cos^2\theta -\mathcal{K}_{-}\sin^2\theta  +\frac{1}{2}\cos 2\theta \sin^2 4 \chi   \right)+\sin 4 \theta -4\theta \right) dV_{S^2}\wedge dV_{dS_2}\,.
\eeq
Finally, the ten-dimensional axion $C_{0}$ is zero while the dilaton is given by
\beq
e^{-\Phi} =\Delta^{-1/2} e^{-2( \alpha+\varphi)} \left( 1+\mathcal{K}_{+} \cos^2\theta \right)\,.
\eeq
We have checked that this solution satisfies the (Lorentzian) equations of motion of type IIB supergravity. 
The fact that the NSNS two-form and the RR four-form potentials are complex is an artifact of the Euclidean nature of our solution.\footnote{By the same token, $EAdS_{5} \times dS_{5}$ is a perfectly good solution of type IIB supergravity, although in that case the four-form potential is also complex.}

\subsection{UV asymptotics of the uplifted solution}
Let us now describe the UV asymptotics of this solution and check that it has the correct falloff expected from the AdS/CFT correspondence. We will again parametrize the radial coordinate as $\rho \equiv e^{-g r}$, so that the boundary is found at $\rho =0$. Starting with the metric we find
\beq
\begin{aligned}
&ds^2_{\text{10D}}\simeq \frac{4}{g^2} \Biggl[ \frac{d\rho^2}{4\rho^2}   + \frac{1}{4\rho} ds^2_{\mathbb{RP}^4} +\cos ^2 \theta  \, d\Omega^{2}_{S^2} +\sin ^2 \theta \, d\Omega^{2}_{dS_2}   \\
&+\rho\, \mathcal{J}^{3/2} \Biggl( \frac{d\rho^2}{4\rho^2} -\frac{2-\mathcal{J}^{3/2}\cos2\theta}{4\,\mathcal{J}^{3/2} \rho}ds^2_{\mathbb{RP}^4} -  \cos ^2 \theta  (\cos 2 \theta +2)d\Omega^{2}_{S^2}- \sin ^2 \theta  (\cos 2 \theta -2)d\Omega^{2}_{dS_2}   \Biggr)  \Biggr]\\
&+\mathcal{O}(\rho^2)
\end{aligned}
\eeq
The first line is the strict UV limit given by the $EAdS_{5}\times dS_{5}$, whereas the second line gives the first subleading term in the radial expansion. From (\ref{bdryexp}), we see that the correction being linear in $\rho$ is consistent with having a boundary operator of dimension $\Delta=2$ that acquires a vev in line with the discussion following the equation  (\ref{bdryexp}).

The expansion of the dilaton near the boundary gives
\beq \label{UVdilaton}
e^{-\Phi} \simeq e^{-2 \varphi_0}+6 \rho^2\, \mathcal{J}^{3/2}  e^{-2 \varphi_0}+\mathcal{O}(\rho^4)
\eeq
From the second term we infer that the dual operator of the dilaton, namely the full $\mathcal{N}=4$ Lagrangian density $\mathcal{L}_{\mathcal{N}=4} \sim \Tr\, F^2+\dots$ acquires a vev. 
From the subleading term, we can determine the corresponding vev after the proper holographic renormalization. 
Similarly, we can obtain the first terms of the expansion of the two-form potentials
\beq
\begin{aligned}
B_{2}& \simeq -\frac{32 i \, \mathcal{J}^{3/2}  e^{\varphi_0} \cos ^3\theta\,\rho ^{3/2} }{g^2} dV_{S^2}+\mathcal{O}(\rho^{5/2}) \\
C_{2}&\simeq \frac{32\,\mathcal{J}^{3/2} e^{\varphi_0}\sin ^3\theta \,\rho ^{3/2} }{g^2} dV_{dS_2}+\mathcal{O}(\rho^{5/2})
\end{aligned}
\eeq
which indicates that there is no source for a dual operator deforming the Lagrangian. On the other hand an operator of dimension $3$ acquires a vev in line with the expectation described before and also with the fact that the lowest Kaluza-Klein mode on $S^5$ of the antisymmetric gauge field is associated to a dimension 3 operator.

Finally, we can determine the UV behaviour of the four-form potential to get
\beq
C_{4}\simeq \frac{2i}{g^4}\left( (\sin 4 \theta -4 \theta )+4\mathcal{J}^{3/2}   \sin ^32 \theta \, \rho \right) dV_{S^2}\wedge dV_{dS_2}+\mathcal{O}(\rho^{2})
\eeq
This expansion is again consistent with (\ref{bdryexp}) and the linear term in $\rho$ contributes to the expectation value of the boundary operator of dimension $\Delta=2$ as expected, but without a source for the Lagrangian deformation at the boundary.

In summary, from the boundary behaviour of the bulk fields we consistently find no deformation of the $\mathcal{N}=4$ SYM Lagrangian but instead some operators acquire vevs, in accordance with field theory expectations.

\subsection{Singular behaviour}
A distinct feature of our solution is the existence of a bulk naked singularity, as was already made clear by the analysis of the five-dimensional scalars.  We now study this singularity in further detail in the uplifted ten-dimensional solution.
The upshot  is that  it exhibits geometric features  reminiscent to those of a O$1_-$ plane. This is somewhat unexpected as 
our background should not contain an orientifold, but of course there is no sharp contradiction. 
In the background (\ref{tendmetric})  one cannot reach the point $r=0$ because the solution becomes unphysical inside a region $r\leq r^{\ast}$ bounded by $r^{\ast}$ precisely given by (\ref{rstar}). So we obtain a  singular behaviour at
\beq \label{singloc}
r^{\ast} \text{ such that }  X(r^{\ast}) =\mathcal{J}\,,\quad \theta^{\ast}=\frac{\pi}{2}\,.
\eeq
Given that the ten dimensional metric (\ref{tendmetric}) is diagonal (see \cite{Bobev:2019wnf} for a general discussion on near-singularity expansions), we can expand each term around (\ref{singloc}) and keep only the leading contribution. We obtain at leading order
\beq
ds^2_{\text{10D}} \simeq  \frac{4}{g^2}\left(3z + \xi^2 \right)^{-3/4}d\Omega_{dS_2}^2+ \frac{4}{g^2} \left(3z + \xi^2 \right)^{1/4} \left(\mathcal{J} dz^2 +\mathcal{J} ds^{2}_{\mathbb{RP}^4} + d\xi^2 +\xi^2 d\Omega_{S^2}^2 \right)
\,,
\eeq
where we have defined 
\beq 
z\equiv  \frac{X(r)-\mathcal{J}}{\mathcal{J} (\mathcal{J}+2)}\quad \rm{and}\quad \xi \equiv \frac{\pi}{2}-\theta.
\eeq
It is clear that at the location specified in (\ref{singloc}) the NSNS two-form potential $B_2$ vanishes while the RR two-form   $C_2$ reads
\beq
C_2 \simeq\,\left(3z + \xi^2 \right)^{-1} e^{-\varphi_0} \sqrt{\frac{\sqrt{\mathcal{J} }+1}{\mathcal{J} +\sqrt{\mathcal{J}  }+1}} \, \frac{4}{g^2} \,dV_{dS_2}\.
\eeq
Finally the behaviour of the ten dimensional dilaton near (\ref{singloc}) is given by
\beq
e^{-\Phi} \simeq e^{-2\varphi_0} \frac{\sqrt{\mathcal{J} }+1}{\mathcal{J} +\sqrt{\mathcal{J} }+1}\, \left(3z + \xi^2 \right)^{-1/2}\,,
\eeq
which again matches (\ref{flat}). 
It turns out that this behaviour precisely resembles a O1$_{-}$ plane in flat space as we now briefly review.

In flat space, the type IIB supergravity solutions corresponding to orientifolds are well known, see for example \cite{Cordova:2019cvf} for a recent review. For the case of a O1$_{-}$ plane, the solution in Einstein's frame is given by the following
\beq \label{flat}
ds^2 = H(r)^{-3/4} ds_{\parallel}^2 +H(r)^{1/4} ds^2_{\perp}\,,\quad e^{\Phi}= H(r)^{1/2}\,,\quad C_{2} = (H(r)^{-1}-1) \rm{vol}_{\parallel}
\eeq
with the harmonic function
\beq
H(r) = 1 - 2^{-4}\frac{(2\pi \ell_s)^{6}g_s}{6\, {\rm{Vol}}_{7}\,r^{6}}\,.
\eeq
In the above solution, $ds_{\parallel}^2$ and $ds^2_{\perp}$ denote the metric on the two parallel directions and  on the eight transverse directions  to the O1 plane respectively, and ${\rm{vol}_{\parallel}}$ is the volume form in the two-dimensional internal manifold. For the O1$_{+}$ solution, the harmonic function $H(r)$ is modified by a sign change in the second term. Importantly, the flat space solution becomes unphysical for the range $r\leq r^{\star}$ where $r^{\star}$ is the point for which the harmonic function vanishes and the metric becomes complex below that value. In addition, the scalar curvature also diverges at $r=r^{\star}$. The region $r\leq r^{\star}$ is the so-called \textit{hole} region, where supergravity cannot be trusted and in particular we cannot reach the location of the orientifold plane at  $r=0$.

In summary, from a purely geometric analysis the singularity (\ref{singloc}) would appear to signal the presence of O1$_{-}$ plane in the bulk geometry, but 
we don't think this is its correct interpretation.
 An orientifold  would involve gauging worldsheet parity, which corresponds to charge conjugation  of the  boundary field theory side, and we are precisely studying the setup that does {\it not} involve  modding out by charge conjugation.
What's more, in the presence of a crosscap the  first correction to the classical background is expected to appear at order $1/N$, in contrast with the diagrammatic analysis 
 section \ref{fieldth}, where we saw that   (in the setup without charge conjugation) it is  of order $1/N^2$.

\subsection{S-duality}\label{secsdual}
From the most general supergravity solution described in appendix \ref{sugradets}, we have encountered two  choices for the boundary value of the five-dimensional scalars $\omega$ and $c$ which are compatible with a vanishing $\theta$ angle. They are
\beq
\omega_0 =0,\, \pi/2\quad {\rm{and}} \quad c_{0}=\pi
\eeq
and as mentioned in the appendix \ref{appuplift}, we could as well set $c_{0}=0$ but that  leads to the same ten-dimensional solution. In the main text we have set $\omega_0 =\pi/2$ but from the boundary field theory point of view, the two solutions are related by the S-duality transformation $\tau \rightarrow -\frac{1}{\tau}$ which preserves $\theta=0$.

We can straightforwardly solve the BPS equations with this new boundary choice $(\omega_0 ,c_0)=(0,0) $ and obtain a new solution. The set of new five-dimensional equations turns out to be the same as before with the only change being a reflection on the five-dimensional dilaton $\varphi \rightarrow -\varphi$. Upon uplifting the resulting solution to ten dimensions we obtain a new set of fields related to the previous choice by
\beq \label{sdual}
   \begin{pmatrix}  \tilde{B}_2 \\
   \tilde{C}_2
  \end{pmatrix} = \begin{pmatrix} 0 & -1   \\
   1
    & 
   0 
   \end{pmatrix}
   \begin{pmatrix}
   B_2 \\
 C_2 
   \end{pmatrix}
\eeq
and
\beq
e^{\tilde\Phi} = e^{-\Phi}, \quad \tilde{C}_0=0\,\quad \tilde{C}_4=C_4\,,
\eeq
which is exactly the S-dual transformed solution. The behaviour close to the singularity is similar to that one of the previous section, with the exception that the NSNS two-form is now non-vanishing and the dilaton is inverted
\beq
\begin{aligned}
\tilde{B}_2 \simeq\,\left(3z + \xi^2 \right)^{-1} e^{-\varphi_0} \sqrt{\frac{\sqrt{\mathcal{J} }+1}{\mathcal{J} +\sqrt{\mathcal{J}  }+1}} \, \frac{4}{g^2} \,dV_{dS_2}\,\\
e^{\tilde{\Phi}} \simeq e^{-2\varphi_0} \frac{\sqrt{\mathcal{J} }+1}{\mathcal{J} +\sqrt{\mathcal{J} }+1}\, \left(3z + \xi^2 \right)^{-1/2}\,,
\end{aligned}
\eeq
as $r\rightarrow r^{\ast}$ and the remaining fields are zero.

\subsection{Completely fixing the solution}
Our solution contains an additional parameter, $\mathcal{J}$, as compared to the standard $AdS_{5}\times S^{5}$ case. Given that $\mathbb{RP}^4$ is non-conformally flat, one expects a coupling to the curvature that cannot be removed for example by a choice of regularization scheme in perturbation theory. For the one-point function of local operators, this coupling might be interpreted as the mixing with the identity operator \cite{Gerchkovitz:2016gxx}. This happens already for observables in the conformally flat sphere $S^4$ but in $\mathbb{RP}^4$ this coupling is physical and not just an artifact of the map to a curved background. Such coupling is only a function of the radius of $\mathbb{RP}^4$ where the boundary theory lives
and once the latter is fixed, there are no further physical parameters besides the Yang-Mills coupling $g_{\text{YM}}$ and the rank of the gauge group $N$. This means that the constant $\mathcal{J}$ is really not a free parameter but rather should be fixed somehow.
In order to do that, we might resort to the computation of some quantity on the gauge theory in the strong coupling regime such as a vev of a certain operator that we can match with supergravity.
Desirably, one could hope for a BPS operator whose vev is protected from quantum corrections, but unfortunately we were not able find such operator. To proceed we are then left with a non-perturbative computation on the field theory side but fortunately we have at our disposal a matrix model \cite{Wang:2020jgh} computing the exact partition function of the theory on $\mathbb{RP}^4$. This matrix model may be solved at least in the large $N$ limit and from which we can read off the strong coupling limit of various quantities. We will report on these results elsewhere \cite{inprogress}.
It would be desirable to find a condition within supergravity which could be used to determine this parameter without appealing to the field theory.

\section{Tree-level one-point functions and integrability} \label{sectree}
As we have emphasized, the new set of observables for $\mathcal{N}=4$ SYM on $\mathbb{RP}^4$ are the vevs of scalar operators. 
In this section, we would like to comment on their computation at tree level. We will use some 
 some familiar tools from integrability (notably the expressions of dilation eigenstates given by the Bethe ansatz) but we'll ultimately find  that in the setup without charge conjugation integrability is broken. By constrast, it is preserved in the setup with charge conjugation, as will be discussed in a separate article~\cite{inprogresscc}.

The simplest example is to consider a subsector involving two complex scalars. In order to have a non-vanishing one-point function each of these complex scalars must involve a combination of two real scalars charged under the two distinct SO(3) factors that compose the R-symmetry\footnote{Here the distinction between SO(3) and SO(2,1) is not important.} group SO(3)$\times$SO(3). Let us define 
\beq
Z\equiv \frac{1}{\sqrt{2}} \left( \Phi_{5} + i\Phi_7 \right)\,,\quad X\equiv \frac{1}{\sqrt{2}} \left( \Phi_{6} + i\Phi_9 \right)\,,
\eeq
where $\Phi_{5,6}$ and $\Phi_{7,9}$ are charged under different SO(3) factors.
Using the propagators (\ref{scalarprop}), this choice leads to the following non-zero contractions as we consider the UV limit 
\beq
\lim_{y\rightarrow x}\, \langle Z(x) Z(y)  \rangle=\lim_{y\rightarrow x}\, \langle X(x) X(y) \rangle  = \frac{g^{2}_{\rm{YM}}} {16 \pi^2} 
\eeq
with the remaining $\lim_{y\rightarrow x} \, \langle Z(x) X(y)\rangle =0 $.

In planar $\mathcal{N}=4$ SYM on flat space, conformal operators are determined by resolving the mixing problem, and at leading order for small 't Hooft coupling their explicit expression can be efficiently handled via the Bethe ansatz. These operators are well described by the Bethe wave-function $\psi$,
\beq
\mathcal{O}_{M}  = \sum_{\vec{n}} \psi(\vec{n},\vec{p} )\, \Tr(Z\dots X \dots X\dots Z) 
\eeq
where $\vec{n}=(n_1,\dots,n_{M})$ are the positions of $X$s inside the trace whose length (i.e. the total number of $X$s and $Z$s) is denoted by $L$ and $\vec{p}$ is the set of Bethe roots satisfying the Bethe equations and vanishing total momentum $\sum_{i}p_i=0$. For example for $M=2$, we have
\beq
\psi(\vec{n},\vec{p})=e^{ip_1 n_1+i p_2 n_2}+S(p_1,p_2) \,e^{ip_1 n_2+i p_2 n_1}\quad{\text{with }}\quad S(p_1,p_2) = -\frac{1+e^{i p_1+i p_2}-2e^{ip_1}}{1+e^{i p_1+i p_2}-2e^{ip_2}}\,,
\eeq
and the set $\vec{p}$ is subject to $e^{i p_{i} L}= \prod_{k\neq i}S(p_i,p_k)$. For the general $M$ case, see for example \cite{Levkovich-Maslyuk:2016kfv} for a recent pedagogical review.

We now consider the one-point functions of such operators in the planar limit, in the setup without charge conjugation. In the absence of $X$s, this amounts to count planar graphs with $L/2$ propagators. It is a well-known fact that this counting problem is solved by the Catalan numbers $C_{n}$ (see for example \cite{Erickson:2000af} for an important application of this number), 
\beq
\langle \Tr\, Z^{L} \rangle = N \left( \frac{g^{2}_{\rm{YM}}N}{16 \pi^2}\right) ^{L/2} C_{L/2}\,,\quad {\text{for even } }L\,,
\eeq
and zero for odd $L$\footnote{We are ignoring a possible additional contribution to this result coming from the non-trivial holonomy of the gauge field along a cycle of $\mathbb{RP}^4$. A more detailed analysis would be needed to determine the effect that this could potentially have even at zero coupling. We thank Shota Komatsu and Kyriakos Papadodimas for comments on this detail.}. As we replace some of the $Z$s by $X$s, the counting of planar diagrams can be determined by the large $N$ limit of the gaussian two-matrix model
\beq \label{omegas}
\omega({\vec{n}})\equiv \langle \Tr(Z\dots X\dots Z)   \rangle =\frac{1}{Z_0}  \int [dZ] [dX]\,  \,\Tr(Z\dots X\dots Z) \, e^{-\frac{N g^{2}}{16 \pi^2} \Tr Z^2-\frac{N g^{2}}{16 \pi^2} \Tr X^2}\,.
\eeq
with $Z_0= \int [dZ] [dX]\,  \, e^{-\frac{N g^{2}}{16 \pi^2} \Tr Z^2-\frac{N g^{2}}{16 \pi^2} \Tr X^2}$.
The one point function is  finally given by
\beq
\langle \mathcal{O}_M \rangle = \sum_{1\leq n_1< \dots<n_{M} \leq L} \psi(\vec{n},\vec{p} ) \,\omega(\vec{n} ) \,.
\eeq
One can in principle determine the numbers $\omega({\vec{n}})$ in full generality from (\ref{omegas}), but for a small number of $X$s it is straightforward to work out the combinatorics directly. For example, for the cases $M=2,4$ (note that $M$ and $L$ have both to be even to obtain a non-vanishing result) we obtain
\beq
\begin{aligned} \label{onept2}
\langle \mathcal{O}_2 \rangle = N\left( \frac{g^{2}_{\rm{YM}}N}{16 \pi^2}\right) ^{L/2}\sum^{}{\vphantom{\sum}}'_{1\leq n_1<n_2\leq L}\,\psi(n_1,n_2)\, C_{\frac{1}{2} (n_1+L-n_2-1)}C_{\frac{1}{2} (n_2-n_1-1)}
\end{aligned}
\eeq
\beq
\begin{aligned} \label{onept4}
\langle \mathcal{O}_4 \rangle = N\left( \frac{g^{2}_{\rm{YM}}N}{16 \pi^2}\right) ^{L/2}\sum_{\vec{n}}^{}{\vphantom{\sum}}'\,&\psi(n_1,n_2,n_3,n_4)\Bigl( C_{\frac{1}{2} (L+n_1-n_2+n_3-n_4-2)}C_{\frac{1}{2} (n_2-n_1-1)}C_{\frac{1}{2} (n_4-n_3-1)}\\
&+C_{\frac{1}{2} (L+n_1-n_4-1)}C_{\frac{1}{2} (n_2-n_3+n_4-n_1-2)}C_{\frac{1}{2} (n_3-n_2-1)} \Bigr)
\end{aligned}
\eeq
where the primed sum $\sum^{}{\vphantom{\sum}}'$ is over the sets $\vec{n}$ for which the indices of all the Catalan numbers are integers. These are not normalized one-point functions so far. We can normalize them by requiring that the UV limit of their two-point functions (i.e. of the operator and its conjugate) has a unit coefficient. This boils down to divide the above one-point functions by the factor
\beq
\mathcal{N} =  \left( \frac{g^{2}_{\rm{YM}}N}{16 \pi^2}\right) ^{L/2} \times \sqrt{\verb+Gaudin norm+}
\eeq
so that they are of order $\mathcal{O}(N)$. In this formula, the Gaudin norm is the norm of the Bethe wave-function, see \cite{Levkovich-Maslyuk:2016kfv} for its explicit expression.

It is natural to inquire whether planar integrability is preserved, in which case one may efficiently
 determine these one-point functions at  the loop level.
To attempt an answer, we can draw some similarities with the defect conformal field theory arising as the dual of the D3-D5 brane system studied in \cite{deLeeuw:2015hxa,Buhl-Mortensen:2015gfd,Buhl-Mortensen:2016pxs,Buhl-Mortensen:2016jqo,Buhl-Mortensen:2017ind,Komatsu:2020sup}  where nontrivial one point functions exist as the result of  a (partial) conformal symmetry breaking. There, the probe D5-brane is described by a boundary state and the one-point function is computed by overlapping it with a closed string state, dual to a single trace operator. Equivalently, the string worldsheet for this defect setup has a disk topology with an insertion of a closed string vertex. 

Despite the differences between the two cases, we might ask if some of the salient features of the weak coupling one-point functions of the defect setup are also present for our case. Notoriously, the D5  boundary is believed to preserve integrability of the original $\mathcal{N}=4$ SYM. A weak coupling signature of this property is the fact that the boundary state is annihilated by the odd spin higher charges out of the infinite set of conserved charges underlying integrability, 
\beq
Q_{2n+1} |\mathcal{B}\rangle =0\,,
\eeq
where $ |\mathcal{B}\rangle$ is the boundary state and $Q_{s}$ denotes the hierarchy of conserved charges in involution with the Hamiltonian $Q_2$.
As a consequence, the overlap with a single trace state is only non-vanishing for the case where the Bethe state is parity-symmetric, or equivalently, the corresponding Bethe rapidities are of the form $\{u_j,-u_j\}_{j=1}^{M/2}$. 

From the explicit results (\ref{onept2}) and  (\ref{onept4}), we can easily check that the same selection rule \textit{does not} hold. In fact, we did not find any selection rule involving other charges or combination thereof, which suggests that this setup does not preserve integrability.

\section{Conclusions} \label{discussion}
In this work we have found a new  background of type IIB supergravity, which we propose as the holographic dual of $\mathcal{N}=4$ SYM theory  on $\mathbb{RP}^4$ in the setup ``without charge conjugation''.
By construction, this solution  respects the same symmetries of the boundary theory and corresponds to a deformation of $\mathcal{N}=4$ SYM by non-trivial one-point functions, as required by the duality.

Our solution contains a bulk naked singularity whose nature is still unclear to us. 
 Geometrically,  the singularity resembles an orientifold  O1$_{-}$ plane, but we do not expect this to be its  correct interpretation,
 as we should not be gauging worlsheet parity (whose field theory counterpart is charge conjugation). The large $N$ expansion of the field theory is also incompatible with the presence of a crosscap in the dual string theory.
On general grounds, we expect this  naked singularity to be resolved within the fully fledged type IIB string theory, as there appears to be nothing singular about the boundary field theory. It would be of great interest to understand how this comes about. A challenge is
that unlike the standard $AdS_5 \times S^5$ background, which  arises as the near-horizon limit of a stack of $D3$ branes in flat space,  we are not aware of an analogous  brane construction for our background. Indeed if one considers a stack of D3 branes with $\mathbb{R}^4$ worldvolume in flat space, the identification that leads to $\mathbb{RP}^4$ is a conformal isometry, which is of course not a symmetry of the full open string field theory --
it  becomes a symmetry only a low-energy, i.e.~in ${\cal N}=4$ SYM. One might consider starting instead with branes with $S^4$ wordvolume, but such a setup is not available in asymptotically flat space.

There are a number of  quantitative checks of the proposed duality that it will be interesting to carry out.  The matrix model of \cite{Wang:2020jgh} can be studied in the large $N$ and large 't Hooft coupling limit, yielding e.g.~the free energy and and the vevs of certain operators. The same observables  can be independently computed in the dual supergravity, taking into account the proper holographic renormalization procedure. An additional difficulty with respect to more standard holographic setups is the presence of the IR naked singularity.
This will be reported elsewhere  \cite{inprogress}.
In particular one should be able to fix the parameter $\mathcal{J}$ that was left free in this solution.

As we have emphasized throughout, the setup studied in this paper is not the unique realization of $\mathcal{N}=4$ on $\mathbb{RP}^4$. There is alternative way, where the spacetime identification of the elementary fields on antipodal points of $S^4$ is  combined with charge conjugation. As described in section \ref{fieldth}, this leads to a very different large $N$ scaling of correlators, with one-point function of order $O(1)$. There is a compelling guess for the holographic dual, as an orientifold projection of $AdS_5\times S^5$, which in particular adds a crosscap on the worldsheet. Recently, crosscap states were studied in the context of two-dimensional integrable field theories \cite{Caetano:2021dbh}, where it was shown that integrability survives in their presence. This suggests that the orientifold setup is in fact integrable and may be studied non-perturbatively in the 't Hooft limit. This is the subject of upcoming work \cite{inprogresscc}.

\section*{Acknowledgments}
The authors are grateful to Shota Komatsu for many inspiring discussions and suggestions throughout this project. We thank Nikolay Bobev,  Fri\dh rik  Gautason and Jesse van Muiden for their critical reading of the manuscript and interesting comments, and
Justin Kaidi,  Kyriakos Papadodimas, Shu-Heng Shao and Yifan Wang for several useful discussions.
The work of L.R.  is supported in part by NSF grant \#~PHY-1915093.

\appendix
\section{Propagators in $\mathbb{RP}^4$} \label{appprops}
In this section, we determine the propagators of the elementary fields in $\mathbb{RP}^4$ seen as $S^{4}$ with the antipodal points identified.
\subsection{Conventions} \label{conventionslag}
We will adopt the following convention for the Euclidean $\mathcal{N}=4$ SYM action (the same as in \cite{Wang:2020jgh}). In $S^{4}$ it reads
\beq 
\begin{aligned} \label{actionsym}
S = -\frac{1}{ 2 g^{2}_{\rm{YM}}} \int_{S^4} d^{4}x\, \tr \Bigl( &\frac{1}{2} F_{\mu \nu}^2 + (D_{\mu} \Phi_{I})^2 - \Psi \Gamma^{\mu} D_{\mu} \Psi+\frac{2}{R^2} \Phi^{I}\Phi_{I} \\
+&   \frac{1}{2} \lbrack \Phi_{I},\Phi_{J} \rbrack \lbrack \Phi^{I},\Phi^{J}\rbrack  - \Psi \Gamma^{I}[\Phi_{I}, \Psi]  \Bigr)
\end{aligned}
\eeq
with the indices $\mu,\nu=1,\dots,4$ and $I,J=5,\dots,9,0$ and $D_{\mu}\, \bullet=\partial_{\mu} + \lbrack A_{\mu},\bullet \rbrack$ and $F_{\mu\nu} =\lbrack D_{\mu},D_{\nu} \rbrack $. The conventions for the SU($N$) generators $T_a$ are such  that $\tr\, T_{a} T_{b} = -\frac{1}{2} \delta_{ab}$ and they are anti-hermitian. 
\subsection{Flat space propagators}
If we consider the theory on $\mathbb{R}^4$, which amounts to the same Lagrangian density as (\ref{actionsym}) except that the conformal masses of the scalars are absent, we have the following propagators
\beq \label{flatprop}
\begin{aligned}
\langle \Phi^{I}(x) \Phi^{J}(y) \rangle_{\mathbb{R}^4} &= \frac{g^{2}_{\rm{YM}}}{4 \pi^2} \frac{\delta^{IJ}}{(x-y)^2}\,, \quad \langle A_{\mu} (x) A_{\nu} (y) \rangle_{\mathbb{R}^4}  = \frac{g^{2}_{\rm{YM}}}{4\pi^2}\frac{\delta_{\mu \nu}}{(x-y)^2} \\
\langle \Psi(x) \Psi(y) \rangle_{\mathbb{R}^4}  &=  - \frac{g^{2}_{\rm{YM}}}{2 \pi^2} \frac{\tilde{\Gamma} \cdot(x-y)}{(x-y)^4}\,,
\end{aligned}
\eeq
where we have used the Feynman gauge for the gauge field propagator and we are omitting the gauge indices. In the short distance limit, the propagators on $\mathbb{RP}^4$ should reduce to these ones.
\subsection{Propagators in $\mathbb{RP}^4$}
We will regard $\mathbb{RP}^4$ as the $\mathbb{Z}_2$ quotient of the sphere $S^4$ and use the stereographic coordinates (\ref{stereo}).
A simple feature that the propagators on $\mathbb{RP}^4$ should reflect is the invariance under the involution $\mathcal{I}$ defined in (\ref{identi}).
For a generic field $\mathcal{\chi}(x)$, one has the relations
\beq \label{invrp4}
\langle \mathcal{\chi}(x) \mathcal{\chi}(y) \rangle_{\mathbb{RP}^4} = 
 \langle  \mathcal{\chi}(x)\mathcal{\chi}'(y') \rangle_{\mathbb{RP}^4}= 
 \langle  \mathcal{\chi}'(x')\mathcal{\chi}(y) \rangle_{\mathbb{RP}^4}= 
 \langle  \mathcal{\chi}'(x')\mathcal{\chi}'(y') \rangle_{\mathbb{RP}^4}
\eeq
where $\mathcal{\chi}'(x') =\mathcal{\chi}(x) $ is the image of $\mathcal{\chi}(x) $ on the covering space $S^4$ specified by the involution $\mathcal{I}$. We will make use of these relations while determining the  propagators. 

\subsubsection{Scalars}
From the equation of motion for the scalar that follows from (\ref{actionsym}), we infer that the propagator obeys
\beq
\left( \nabla_{S^4}^2-\frac{2}{R^2}\right)G_{\Phi}(x,y) =0\,,
\eeq
for $x\neq y$ and the scalar propagator is related to $G_{\Phi}(x,y)$ by
\beq 
\langle \Phi^{I}(x)\Phi^{J}(y)\rangle= \delta^{IJ}G_{\Phi}(x,y)\,.
\eeq
This equation is solved by
\beq
G_{\Phi}(x,y) = \frac{a}{\eta} +\frac{b}{1-\eta}
\eeq
for any two constants $a,b$ and $\eta$ is the chordal distance on the sphere given by
\beq
\eta = \frac{(x-y)^2}{(1+x^2)(1+y^2)}\,.
\eeq
We impose the conditions (\ref{invrp4}) using that under the involution $\mathcal{I}$ the image of the scalar is
\beq
\Phi^{I} (x) =  (\mathcal{R} \Phi)^I (-x^{\mu}/|x|^2 ) \equiv (\Phi')^{I} (x')\,.
\eeq
Finally, one can fix the singular behaviour  in the limit as $y\rightarrow x$ by matching with flat space (\ref{flatprop}) to finally obtain
\beq \label{scalarprop}
\langle \Phi^{I} (x)  \Phi^{J} (x)  \rangle_{\mathbb{RP}^4} =  \frac{1}{4} \frac{g^{2}_{\rm{YM}} \delta^{IJ}}{4 \pi^2} \left(\frac{1}{\eta} \pm \frac{1}{1-\eta}\right)\,.
\eeq
where the sign depends on the parity of the field under $\mathcal{R}$, namely plus or minus for $I=4,5,6$ or $I=1,2,3$ respectively. This propagator has also been determined in \cite{Giombi:2020xah}.

\subsubsection{Fermions}
Using analogous procedure for the fermions, we can check that
\beq \label{fermiprop}
G_{\Psi}(x,y)\equiv \langle \Psi(x) \Psi(y) \rangle_{\mathbb{RP}^{4}} = -\frac{g^{2}_{\rm{YM}}}{2 \pi^2}\frac{1}{ 16}\left(\frac{\tilde{\Gamma} \cdot(x-y)}{\eta^2}+\frac{ i}{(1-\eta)^2}\widetilde{\mathcal{R}}(1+\Gamma \cdot x\, \widetilde{\Gamma}\cdot y) \right)\,.
\eeq
satisfies the equation for the fermionic propagators
\beq
\Gamma^{\mu}  D_{\mu} \,  G_{\Psi}(x,y) = 0\,,
\eeq
for $x\neq y$. In checking this equation, it may be useful  to use the following properties of $\mathcal{R}$,
\beq
\mathcal{R} \widetilde{\Gamma}_{\mu} = - \Gamma_{\mu} \widetilde{\mathcal{R}} \,, \quad \mathcal{R}\widetilde{\mathcal{R}} =-1\,,
\eeq
where $\widetilde{\mathcal{R}} =  -\widetilde{\Gamma}_{[7} \Gamma_{9} \widetilde{\Gamma}_{0]}$. Moreover, this propagator satisfies (\ref{invrp4}) where we use that the image of the fermion under $\mathcal{I}$ is given by
\beq
\Psi(x) = \frac{i}{|x|} \widetilde{\Gamma}_{\mu} x^{\mu} \mathcal{R} \Psi(-x^{\mu}/x^2) \equiv \Psi'(x')\,,
\eeq
and the singularity  matches the flat space one (\ref{flatprop}) in the limit where $y\rightarrow x$.

\subsubsection{Gauge Fields}
Here we adapt the method of \cite{DHoker:1999bve} to determine the gauge propagator. The idea is to start by writing the bi-tensor $G_{\mu \nu}(\eta)\equiv\langle A_{\mu} (x) A_{\nu} (y) \rangle $  as
\beq \label{propgauge}
G_{\mu \nu}(\eta)
= -\left(\partial_{\mu}^{x} \partial_{\nu}^{y} \eta \right) F(\eta) + \partial^{x}_{\mu} \partial^{y}_{\nu} S(\eta)\,.
\eeq
where we have ignored color indices.
From now on we will discard $S(\eta)$ since it drops from the equation of motion and will now determine $F(\eta)$. 
The following properties of the derivatives of $\eta$ will be useful,
\beq
\begin{aligned}
D^{\mu} \partial_{\mu}\eta &=2 (1-2\eta)\\
D^{\mu}\eta \partial_{\mu} \eta &=\eta(1-\eta) \\
D_{\mu}\partial_{\nu}\eta &= \frac{1}{2} g_{\mu \nu} (1-2\eta)\\
(\partial_{\mu} \eta)(D^{\mu} \partial_{\nu} \partial_{\nu'} \eta) &= - \partial_{\nu}\eta  \partial_{\nu'} \eta \\
D_{\mu}\partial_{\nu}\partial_{\nu'}\eta &=- g_{\mu \nu} \partial_{\nu'}\eta  \\
D^{\mu} \eta \partial_{\mu} \partial_{\nu '} \eta &= \frac{1}{2}(1-2\eta)\partial_{\nu'}\eta
\end{aligned}
\eeq
where primed derivatives are taken with respect to $y$. The propagator satisfies the following equation
\beq \label{eomgauge}
D^{\mu} \partial_{\mu} G_{\nu \nu'} -D^{\mu} \partial_{\nu} G_{\mu \nu'} = -g_{\nu \nu'} \delta(x,y) + \partial_{\nu'} \Lambda_{\nu} (x,y)
\eeq
and for $x\neq y$ we ignore the delta function. Plugging (\ref{propgauge}) (ignoring the second term) in  the LHS of (\ref{eomgauge}) and using the above properties we get the following equation
\beq
\begin{aligned}
D^{\mu} \partial_{\mu} G_{\nu \nu'} -D^{\mu} \partial_{\nu} G_{\mu \nu'}  &= \partial_{\nu}\eta \partial_{\nu '} \eta \left(-3 F'+\frac{1}{2}(1-2\eta) F''\right)\\
&+\partial_{\nu} \partial_{\nu'} \eta \left(-\frac{3}{2}(1-2\eta) F' -\eta(1-\eta) F''\right)\,.
\end{aligned}
\eeq
Using that $\Lambda_{\nu} = \partial_{\nu} \eta\, \Lambda(\eta)$ we finally get the equations
\beq
\begin{aligned} \label{eqsf}
-3 F'+\frac{1}{2}(1-2\eta) F''&= \Lambda'\\
-\frac{3}{2}(1-2\eta) F' -\eta(1-\eta) F''&=\Lambda
\end{aligned}
\eeq
We integrate the first equation
\beq
\Lambda = -2 F +\frac{1}{2}(1-2 \eta) F' + \Lambda_0
\eeq
where $\Lambda_0$ is an integration constant. We plug this in the second equation of (\ref{eqsf}) and get
\beq
\eta (\eta-1) F'' +2(2\eta-1) F' +2 F -\Lambda_0=0
\eeq
whose general solution is given by
\beq
F(\eta) =\frac{ \Lambda_0}{2} + \frac{a}{\eta} +\frac{b}{1-\eta}
\eeq
for arbitrary constants $a$ and $b$.
We set $\Lambda_0=0$ so that the correlation vanishes for large separation. The result is then
\beq
\langle A_{\mu}(x) A_{\nu} (y) \rangle = - \partial_{\mu} \partial_{\nu} \eta \left( \frac{a}{\eta} +\frac{b}{1-\eta}   \right)
\eeq
To fix these constants, we use that the correlator has to be invariant under the involution $\mathcal{I}$ according to (\ref{invrp4}). For example, using
\beq
\langle A_{\mu} (x) A_{\nu}(y) \rangle =- \langle A_{\mu} (x) I_{\nu}{}^{\rho}(y) A_{\rho}(y') \rangle\,,
\eeq
we obtain the restriction
\beq
b=-a
\eeq
and the remaining constant is fixed by normalization, namely for $\eta \rightarrow 0$, one matches the flat space propagator (\ref{flatprop}). We finally get 
\beq
\langle A_{\mu}(x) A_{\nu} (y) \rangle =- \frac{1}{4 \pi^2} \partial_{\mu} \partial_{\nu} \eta\, \frac{g^{2}_{\rm{YM}}}{2} \left( \frac{1}{\eta} -\frac{1}{1-\eta}   \right)\,.
\eeq

\section{General five-scalar solution} \label{sugradets}
In this section, we construct the most general solution out of the five-scalar truncation (\ref{trunc5}) that satisfies our ansatz.
We will solve the  five-dimensional BPS equations and then check that the resulting solution uplifts to a ten-dimensional solution of type IIB supergravity.

The BPS equations arise as conditions for the vanishing of the supersymmetric variations of the spin 1/2 and spin 3/2 fields. To derive such equations in the five-dimensional $\mathcal{N}=8$ supergravity, the procedure is systematic
and thoroughly described in \cite{Gunaydin:1985cu}. Since we are using the same  27-bein of E$_{6(6)}$ parametrizing the scalar coset (\ref{coset}) as in \cite{Bobev:2020fon}  (see also  our appendix \ref{appuplift} for its explicit expression), many equations can be recycled from there. We refer the reader to  appendix A of~\cite{Bobev:2020fon}  for further details on the derivation.

 The main difference here arises from the analytic continuation to Euclidean signature. After this continuation, most of the equations are written in terms of the superpotentials $\mathcal{W}$ and $\widetilde{\mathcal{W}}$ given by
\beq
\begin{aligned} \label{superpots}
\mathcal{W} &= -\frac{3g}{2}  (\cosh 2 \alpha  \cos 2 \chi+\sinh 2 \alpha\,  \sin 2 \chi  )\,,\\
\widetilde{\mathcal{W}} &= -\frac{3g}{2}  (\cosh 2 \alpha  \cos 2 \chi -\sinh 2 \alpha\,  \sin 2 \chi )
\end{aligned}
\eeq
In particular, notice that $\widetilde{\mathcal{W}}$ is not the complex conjugate of $\mathcal{W}$ as opposed to the Lorentzian solution in \cite{Bobev:2020fon}.

\paragraph{Spin-$1/2$ equations.} In terms of these superpotentials, the spin-1/2 equations can be written as
\beq
i\beta\, \gamma_{r} \,\epsilon^{s} = \frac{1+\sqrt{2}}{6} \partial_{\alpha}\widetilde{\mathcal{W}}\, \epsilon^{\bar{s}}
\eeq
where $\beta$ is a combination of scalars that will appear often $\beta \equiv \alpha' -\sec(c+2 \omega)\varphi'$ with prime denoting derivatives with respect to $r$, $\gamma_{r}$ is the five-dimensional gamma matrix along the $r$ direction and $\epsilon^{s,\bar{s}=1,2,3,4}$ are the spinors parametrizing the supercharges. Given that each spinor $\epsilon^{s}$ has four real independent components by the symplectic Majorana condition and $\epsilon^{\bar{s}}$ is completely determined by $\epsilon^s$, this gives the expected sixteen independent real supercharges. The remaining spin-1/2 equations read
\beq
\begin{aligned}
\beta^2 &= \frac{1}{36} \partial_{\alpha} \mathcal{W}\, \partial_{\alpha} \widetilde{\mathcal{W}}\\
\chi' \, \beta &= \frac{1}{24} \sin 4\chi \left( \frac{\mathcal{W}\partial_{\alpha} \mathcal{W} +\widetilde{\mathcal{W}} \partial_{\alpha}\widetilde{\mathcal{W}}}{2} \right) \\
(\alpha' -\beta)\beta &=\frac{1}{24} \tan 4\chi \left( \frac{\mathcal{W}\partial_{\alpha} \mathcal{W}-\widetilde{\mathcal{W}} \partial_{\alpha}\widetilde{\mathcal{W}}}{2}\right)
\end{aligned}
\eeq
and finally
\beq
\begin{aligned} \label{eqsoc}
\omega' &=\sinh^2 \varphi\, c' \\
c' &= -\frac{2 \tan (c+2\omega) \varphi'}{\sinh 2 \varphi}\,.
\end{aligned}
\eeq
Equations (\ref{eqsoc}) imply that the combination
\beq
\mathcal{C} \equiv \sinh 2\varphi \sin (c+2 \omega)
\eeq
is conserved. As described in the main text, the scalars $c$ and $\omega$ are dual to the $\theta$ angle of Super Yang-Mills and to phase of the operator of dimension three discussed around (\ref{dualops}), respectively. These parameters are fixed on the field theory side. The $\theta$ angle of Yang-Mills on $\mathbb{RP}^4$ can take up only two values, $\theta=0\,,\pi$.  Throughout this paper we focus on the case $\theta=0$. The theta angle is precisely identified with the boundary value of the ten-dimensional axion. The uplift of the five dimensional axion $c$ to the corresponding ten dimensional one will be worked out in the appendix \ref{appuplift}, but we anticipate that by setting the boundary value of $c(r\rightarrow \infty)$ to $\pi$ forces the corresponding boundary ten dimensional axion to be zero as well.

The boundary value of the phase $\omega$ follows from CPT invariance together with the previous requirement that the theta angle is fixed to be zero. In flat space, the phase of  the  bilinear fermionic operator $\tr ( \lambda \lambda)$ (omitting indices and the projection into the SO(3)$\times$SO(3) singlet)  in the $\bf{10}$ of SU(4) can be modified by an SL(2,$\mathbb{Z}$) transformation. Since this operator can be obtained by the action of two charges $Q$ on a $\bf{20}'$ lowest weight, which is invariant under  SL(2,$\mathbb{Z}$), we get that the operator must transform as a modular form of weights $(1/2,-1/2)$ since each $Q$ itself transforms with weights $(1/4,-1/4)$ \cite{Intriligator:1998ig}, namely
\beq
\tr ( \lambda \lambda) \xmapsto{SL(2,\mathbb{Z})} (c \tau + d )^{1/2}(c \bar{\tau} + d )^{-1/2}\, \tr ( \lambda \lambda)  \equiv e^{i \xi }\, \tr ( \lambda \lambda) 
\eeq 
and the complex couplings transform as $\tau \mapsto \frac{a \tau + b}{c\tau +d}$ and $\bar\tau \mapsto \frac{a \bar\tau + b}{c \bar\tau +d}$ for $ \bigl( \begin{smallmatrix}a & b\\ c & d\end{smallmatrix}\bigr)\in {\rm{SL}}(2,\mathbb{Z})$ and $\xi = \arg(c\tau +d)$. For the operator $\tr ( \bar\lambda \bar\lambda)$ in the $\bar{\bf{10}}$, the modular weights are instead  $(-1/2,1/2)$.
 
In $\mathbb{RP}^4$ however, the allowed  SL(2,$\mathbb{Z}$) transformations should preserve the condition $\theta=0$ which implies that the parameter $a,b,c,d$ must obey either
\beq \label{solsl2z}
b=c=0\,,\;\; d=1/a\quad {\rm{or}}\quad a=d=0\,,\;\; b=-1/c
\eeq
which sets  $\xi=0,\pi$ or $\xi=\pm \pi/2$ respectively. For each solution in (\ref{solsl2z}), the two choices of $\xi$ lead to an equivalent type IIB solution so we take one representative, namely $\xi=0$ or $\xi=\pi/2$.  We see that the phase of the operator cannot be further modified by a modular transformation and we can determine it from CPT invariance.

After the analytic continuation to imaginary time, CPT in Euclidean signature amounts to a  $\pi$-rotation on the plane defined by one space coordinate and the analytically continued time direction, besides the charge conjugation. 
For the one point function of the bilinear fermionic operator, CPT acts as 
\beq
\text{CPT}: \tr \,(\lambda \lambda) (x) \mapsto - \tr\, ({\bar{\lambda}}  {\bar{\lambda}}) (\tilde{x})
\eeq
where the sign comes from a $(-1)^{1/2}$ factor for each fermion under CPT, $\tilde{x}$ represents the $\pi$-rotation in a plane formed by two coordinates of $x^{\mu}$. From the CPT invariance of the theory, we expect the one-point function of both these operators to be the same.   
Their modulus is the same and controlled by the asymptotic value of the scalar $\chi$ whereas the corresponding phases are determined by the asymptotic value of the scalar $\omega$. If
\beq
\langle \tr \,(\lambda \lambda) (x) \rangle = \rho \, e^{i\xi +i\phi}
\eeq
then we have $\langle \tr \,(\bar\lambda \bar\lambda) (\tilde{x}) \rangle = \rho \,e^{-i \phi-i \xi}$, given that the latter carries opposite U(1)$\subset$ SL(2,$\mathbb{R}$) charge (note that the one point function does not depend on the insertion of the operator, so $\rho$ and $\phi$ are independent of the position). Therefore we get that $ \rho \, e^{i \phi+i \xi}=- \rho \, e^{-i \phi-i\xi}$ which gives $\phi = \pm \pi/2$ or $\phi=0,\pi$ for the two choices of $\xi$ respectively. We note again that for a given  value of $\xi$ the two resulting possibilities for $\phi$ are physically equivalent, so  we  may set $\phi=0$ or $\phi=\pi/2$. For the main part of the paper, we have set $\phi = \pi/2$ and we comment on the other choice in relation to the S-duality in  section \ref{secsdual}.

We can verify that this condition on the phase of the fermionic bilinear operator is consistent with the identification (\ref{identi}). The map of the ten-dimensional fermions under the involution translates into the following transformation of the corresponding four-dimensional Weyl fermions
\beq
\lambda_{\alpha a \dot{a}} = - i \frac{\sigma_{\mu} x^{\mu}}{|x|} \left(\sigma^{1}\right)^{b}{}_{a} \, \bar{\lambda}_{\dot{\alpha} b \dot{a}}
\eeq
where $\sigma^1$ is the Pauli matrix, $\sigma^{\mu}_{\alpha \dot{\alpha}} = (\vec{\sigma},-i)$ and  we have split the $\bf{4}$ indices of SU(4) into a pair of fundamental SU(2) indices, $A \mapsto (a,\dot{a})$. This transformation implies the relation between the one-point functions of the above operators
\beq
\langle  \tr \,(\lambda \lambda) (x) \rangle = -\langle  \tr \,(\bar{\lambda} \bar{\lambda}) (x') \rangle
\eeq
where $(x')^{\mu} \equiv -\frac{x^{\mu}}{x^2}$, from which we obtain the same constraint on the bulk scalar dual to the phase of these operators.

We can verify this result at tree level in the gauge theory by computing the Wick contraction using the propagator (\ref{fermiprop}). We get such phase from the factor of $i$ in front of the second term of the propagator which is the only contribution to the one-point function as one takes the limit of coinciding points $y \rightarrow x$.

This fixes the value $\omega(r\rightarrow \infty)=\pm \pi/2$ (we can choose the plus sign, given that the relevant phase is $2\omega$). Together with the above condition $c(r\rightarrow \infty)=\pi$ we determine the constant $\mathcal{C}=0$. We then consistently set these two scalars to constants in the bulk with their value being the same as the boundary ones. This last step, although consistent with the equations of motion, is not strictly necessary, but we take it as part of our ansatz. 

\paragraph{Spin-$3/2$ equations.} The spin-3/2 equations can again be obtained by performing the analytical continuation (\ref{ancont}) to the Euclidean signature on the corresponding ones in \cite{Bobev:2020fon}. They look as follows
\beq
  \nabla_{i}  \begin{pmatrix}  \epsilon^{s} \\
  \tilde  \epsilon^{\bar{s}} 
  \end{pmatrix} = \begin{pmatrix} 0 &
   \frac{i(1+\sqrt{2})}{6}\widetilde{\mathcal{W}}   \\
    -\frac{i(-1+\sqrt{2})}{6} \mathcal{W}
    & 
   0 
   \end{pmatrix}
   \gamma_i  
   \begin{pmatrix}
    \epsilon^{s} \\
 \epsilon^{\bar{s}} 
   \end{pmatrix}
\eeq
where $i$ is an index on the $S^4$ slice\footnote{At this point, the distinction between  $\mathbb{RP}^4$ or its double cover $S^4$ is not relevant.}, see (\ref{metric}). There is also an additional equation fixing the radial dependence of the spinor but that will not be important for our purposes.
We can rewrite the covariant derivative in terms of a derivative on the unit-radius sphere
\beq
\nabla_{i}=\widehat{\nabla}_{\hat \imath}-\frac{1}{2} A' \gamma_{r} \gamma_{i}
\eeq
with $\hat \imath$ being the index on the unit $S^4$. The spinors on the unit $S^4$ satisfy
\beq
\widehat{\nabla}_{\hat \imath} \epsilon =  \pm \frac{1}{2} \gamma_{5}\gamma_{\hat\imath} \epsilon
\eeq
where $\gamma_{5} = \gamma_{\hat{0}\hat{1}\hat{2}\hat{3}} =\gamma_{r} $, and all gamma matrices $\gamma_{\hat{\imath}}$ are Euclidean. Note that  $\gamma_{\hat\imath} =e^{-A} \gamma_{i} $. 
Using the Killing spinor equation on $S^4$ we arrive at
\beq \label{spin32}
  \begin{pmatrix} \kappa\, e^{-A}   + A' & 0
    \\
    0
    & 
   \kappa\, e^{-A}   + A'  
   \end{pmatrix}  \gamma_{r} \begin{pmatrix}  \epsilon^{s} \\
  \tilde  \epsilon^{\bar{s}} 
  \end{pmatrix} = \begin{pmatrix} 0 & \frac{i(1+\sqrt{2})}{3} \widetilde{\mathcal{W}}   \\
    -\frac{i(-1+\sqrt{2})}{3} \mathcal{W}
    & 
   0 
   \end{pmatrix}
   \begin{pmatrix}
    \epsilon^{s} \\
 \epsilon^{\bar{s}} 
   \end{pmatrix}
\eeq

These equations are obeyed for any $\epsilon^{s,\bar{s}}$ provided the following conditions are satisfied
\beq \label{grav0e}
(- e^{-A} \kappa  + A' )(- e^{-A} \widetilde\kappa  + A' ) = \frac{\mathcal{W} \widetilde{\mathcal{W}}}{9}\,.
\eeq
with $\kappa,\widetilde{\kappa}=\pm1$.
If we turn off the scalars, $\mathcal{W} =\widetilde{\mathcal{W}} = -3 g/2$, and in addition set $\widetilde{\kappa}=-\kappa$ we recover pure $AdS_5$. From now on, we will take $\widetilde{\kappa}=-\kappa=1$.

\section{Details of the ten-dimensional uplift to type IIB} \label{appuplift}
We provide some details on the uplift of the five-dimensional solution to the type IIB supergravity. The formulae in this appendix has been derived in \cite{Baguet:2015sma,Lee:2014mla} and already applied in several instances such as \cite{Bobev:2018hbq, Bobev:2018eer,Petrini:2018pjk,Bobev:2018ugk,Bobev:2020fon}. 
The uplift procedure amounts to  reconstruct in a consistent way the additional five coordinates completing the ten-dimensional background. For the vacuum solution, i.e. with no additional fields turned on besides the metric, these additional five-dimensional space uplifts to the sphere $S^5$. In Euclidean signature, we need to additionally continue the sphere to the five-dimensional de sitter $dS_5$. We recycle the Lorentzian formulae of \cite{Bobev:2018hbq} by complexifying  one of the embedding coordinates. Given the residual R-symmetry preserved by the boundary theory, it is natural to consider the embedding on $dS_5$ as follows
\beq \label{embup}
\begin{aligned}
&Y_{1} = \cos \theta \cos \phi_1\,, \quad Y_{2} = \cos \theta \sin \phi_1 \cos \phi_2\,, \quad Y_{3} = \cos \theta \sin \phi_1 \sin \phi_2 \\
&Y_{4} =-i  \sin \theta \sinh \chi_1\,, \quad Y_{5} = \sin \theta \cosh \chi_1 \cos \chi_2\,, \quad Y_{6} = \sin \theta \cosh \chi_1 \sin \chi_2
\end{aligned}
\eeq
where we have analytically continued $Y_{4}$. 

The next ingredient for the uplift is to choose a representative element $U$ of  E$_{6(6)}$ parametrizing the scalar coset (\ref{coset}). We make a choice analogous to the one in \cite{Bobev:2020fon}, which we will explicit show here for completeness. We start with a basis for $\mathfrak{e}_{6(6)}$ parametrized by the element $X$ which is a 27$\times$27 matrix with the following block structure
\beq
X =\left[
\begin{array}{c;{2pt/2pt}c}
 A_{15\times15} & B_{15\times12} \\ \hdashline[2pt/2pt]
 C_{12\times15} & D_{12\times12} \\
\end{array}
\right]\,.
\eeq
The non-zero entries are easily determined by finding the generators of $\mathfrak{e}_{6(6)}$ which preserve $\mathfrak{so}(3) \times \mathfrak{so}(3) \subset \mathfrak{so}(6)$.
The axio-dilaton submanifold SL(2,$\mathbb{R}$)/SO(2) is generated by two generators
\beq
\mathfrak{r} =\left[
\begin{array}{c;{2pt/2pt}c}
{\bf{0}}_{15\times15} & {\bf{0}}_{15\times12} \\ \hdashline[2pt/2pt]
{\bf{0}}_{12\times15} & {\bf{1}}_{6\times6} \otimes ( -i\sigma_2)\\
\end{array}
\right]\,,\quad 
\mathfrak{t} =\left[
\begin{array}{c;{2pt/2pt}c}
{\bf{0}}_{15\times15} & {\bf{0}}_{15\times12} \\ \hdashline[2pt/2pt]
{\bf{0}}_{12\times15} & {\bf{1}}_{6\times6} \otimes \sigma_1\\
\end{array}
\right] \,,
\eeq
with $\sigma_{i}$ being the standard Pauli matrices. Then this coset is parametrized by
\beq
U_{\varphi}=e^{-c\, \mathfrak{r}/2} \, e^{-\varphi\, \mathfrak{t}}\, e^{(c/2+\pi/4) \mathfrak{r}} 
\eeq
and note that in our case\footnote{Our parametrization differs by the one used in \cite{Bobev:2020fon} by $\varphi \rightarrow -\varphi$.} we will set $c(r)=\pi$. The remaining part of the coset (\ref{coset}) involves  the generators
\beq
\mathfrak{g}_{\chi} =\left[
\begin{array}{c;{2pt/2pt}c}
{\bf{0}}_{15\times15} & {\mathcal{M}} \\ \hdashline[2pt/2pt]
{\mathcal{M}^\intercal} & {\bf{0}}_{12\times12}\\
\end{array}
\right]\,,\quad    \mathfrak{g}_{\alpha} =\left[
\begin{array}{c;{2pt/2pt}c}
{\mathcal{N}} & {\bf{0}}_{15\times12} \\ \hdashline[2pt/2pt]
{\bf{0}}_{12\times15} & \sigma_{2}\otimes {\bf{1}}_{6\times6}\\
\end{array}
\right]
\eeq
where the matrices $\mathcal{M}$ and $\mathcal{N}$ are
\beq
\mathcal{M}=\scriptsize{\left[
\begin{array}{cccccc;{2pt/2pt}cccccc}
 0 & 0 & 0 & 0 & \sqrt{2} & \sqrt{2} &\\
 0 & 0 & -\sqrt{2} & -\sqrt{2} & 0 & 0 &  \\
 0 & 0 & 0 & 0 & 0 & 0 &  \\
 0 & 0 & 0 & 0 & 0 & 0 &  \\
 0 & 0 & 0 & 0 & 0 & 0 &  \\
 \sqrt{2} & \sqrt{2} & 0 & 0 & 0 & 0 & \\ \hdashline[2pt/2pt]
   &  & &   &  &  & && && \\ \hdashline[2pt/2pt]
  &  &  &  & &  & 0 & 0 & 0 & 0 & -\sqrt{2} & \sqrt{2} \\
 &  &  &  &  &  & 0 & 0 & \sqrt{2} & -\sqrt{2} & 0 & 0 \\
  &  &  &  &  &  & -\sqrt{2} & \sqrt{2} & 0 & 0 & 0 & 0 \\
\end{array}
\right]},\quad
\mathcal{N}=\scriptsize{\left[
\begin{array}{cccccc;{2pt/2pt}cccccc;{2pt/2pt}ccc}
 -2 & 0 & 0 & 0 & 0 & 0 &  &  &  &  &  &  &  &  &  \\
 0 & -2 & 0 & 0 & 0 & 0 &  &  &  &  &  &  &  &  & \\
 0 & 0 & 0 & 0 & 0 & 0 &  &  &  &  &  &  &  &  &\\
 0 & 0 & 0 & 0 & 0 & 0 &  &  &  &  &  &  &  &  & \\
 0 & 0 & 0 & 0 & 0 & 0 &   &  &  &  &  &  &  &  &\\
 0 & 0 & 0 & 0 & 0 & -2 & &  &  &  &  &  &  &  &\\ \hdashline[2pt/2pt]
    &  &  &  & &  & &  &  &  &  &  &  &  &  \\
 \hdashline[2pt/2pt]
  & &  & & &  &   &  &  &  &  &  & 2 & 0 & 0 \\
  &  &  &  & &  &  &  &  &  &  &  & 0 & 2 & 0 \\
  &  & &  &  &  &  &  &  & &  &  & 0 & 0 & 2 \\
\end{array}
\right]}
\eeq
such that the final parametrization of  (\ref{coset}) is given by
\beq \label{paramU}
U= e^{i\chi\,  \mathfrak{g}_{\chi}}e^{\alpha\, \mathfrak{g}_{\alpha}} e^{- \omega \, \mathfrak{r} }\,U_{\varphi}
\eeq
and we end up setting $\omega(r) = \pi/2$. Once the element $U$ and the embedding coordinates (\ref{embup}) are fixed, it is a systematic procedure to apply the formulae \cite{Baguet:2015sma} which is nicely summarized in \cite{Bobev:2018hbq} to generate the ten-dimensional uplifted solution. 

Importantly for our particular solution is the explicit expression for the ten dimensional axion. From the uplift formulae we obtain
\beq
C_0= \frac{e^{2 \alpha} ( \mathcal{K}_{-} \sin^2 \theta+1)
\mathcal{L}_{c}^{+}\mathcal{L}_{s}^{-}-e^{-2 \alpha} (\mathcal{K}_{+} \cos^2 \theta+1) \mathcal{L}_c^{-} \mathcal{L}_s^{+}}{e^{-2\alpha} ( \mathcal{K}_{+}\cos^2 \theta +1)( \mathcal{L}_c^{-})^2+e^{2\alpha} (\mathcal{K}_{-} \sin^2 \theta +1)( \mathcal{L}_s^{-})^2}
\eeq
where $\mathcal{K}_{\pm}$ was defined below equation (\ref{tendmetric}) and $\mathcal{L}_{c,s}^{\pm}$ are defined as
\beq
\mathcal{L}_{c}^{\pm} = \cosh (\varphi ) \cos (\omega )\pm \sinh (\varphi) \cos (c+\omega )\,,\quad  \mathcal{L}_{s}^{\pm} = \cosh (\varphi ) \sin (\omega )\pm\sinh (\varphi ) \sin (c+\omega)
\eeq
One can easily determine that for $r\rightarrow \infty$ the axion becomes
\beq
C_0 \rightarrow \frac{2 \sinh (\varphi_{0}) \cosh (\varphi_{0}) \sin (c_{0})}{\sinh (2 \varphi_0) \cos (c_0)-\cosh (2 \varphi_0)}
\eeq
where $c_0$ and $\varphi_0$ are the asymptotic values of $c$ and $\varphi$. We see that choosing $c_0 = 0$ or $\pi$ gives $C_0=0$ consistently with the value of the $\theta$ angle in the field theory. Both choices of $c_{0}$ lead to the same physical solution and in this paper we set $c_0 = \pi$.

\pdfbookmark[1]{\refname}{references}
\bibliographystyle{JHEP}
\bibliography{main}

\end{document}